\documentclass[submission, PhysComRep]{SciPost_better_arXiv}
 \usepackage{xspace}
 \usepackage{dcolumn}
 \usepackage{mdframed}
 \usepackage{tabularx}
 \hypersetup{
    colorlinks,
    linkcolor={red!50!black},
    citecolor={blue!50!black},
    urlcolor={blue!80!black}
}

 \usepackage[bitstream-charter]{mathdesign}
 \usepackage[normalem]{ulem}
 \usepackage{graphicx}
 \usepackage{subcaption}

\definecolor{comment}{rgb}{0,0.3,0}
\definecolor{identifier}{rgb}{0.0,0,0.3}

\usepackage{booktabs}
\DeclareSymbolFont{usualmathcal}{OMS}{cmsy}{m}{n}
\DeclareSymbolFontAlphabet{\mathcal}{usualmathcal}

\let\oldhref\href
\renewcommand{\href}[2]{\oldhref{#1}{\hbox{#2}}}

\definecolor{lightblue}{rgb}{0.0,0.5,1.0}
\definecolor{darkgreen}{rgb}{0.0, 0.5, 0.0} 

\newcommand{\pb}{{\ensuremath\unskip\,\text{pb}}\xspace}
\def\refeq#1{\mbox{(\ref{#1})}}

\def\citere#1{\mbox{Ref.~\cite{#1}}}

\newcommand{\ie}{\emph{i.e.}\ }

\def\be{\begin{equation}}
\def\ee{\end{equation}}

\newcommand{\PH}{\ensuremath{\text{H}}\xspace}
\newcommand{\Pj}{\ensuremath{\text{j}}\xspace}
\newcommand{\Pp}{\ensuremath{\text{p}}\xspace}

\newcommand{\Pb}{\ensuremath{\text{b}}\xspace}

\newcommand{\Pt}{\ensuremath{\text{t}}\xspace}

\newcommand{\Pg}{\ensuremath{\text{g}}\xspace}

\newcommand{\PW}{\ensuremath{\text{W}}\xspace}
\newcommand{\PZ}{\ensuremath{\text{Z}}\xspace}

\newcommand{\process}{\Pp\Pp\to\PH\Pj\Pj}

\newcommand{\processfss}{\PH\Pj\Pj}

\newcommand{\mjj}{m_{\Pj\Pj}}
                                    
\newcommand{\Mt}{\ensuremath{m_\Pt}\xspace}

\newcommand{\MW}{\ensuremath{M_\PW}\xspace}

\newcommand{\MZ}{\ensuremath{M_\PZ}\xspace}
\newcommand{\Mb}{\ensuremath{m_\Pb}\xspace}

\newcommand{\GH}{\ensuremath{\Gamma_\PH}\xspace}
\newcommand{\GZ}{\ensuremath{\Gamma_\PZ}\xspace}

\newcommand{\GW}{\ensuremath{\Gamma_\PW}\xspace}

\newcommand{\GeV}{\ensuremath{\,\text{GeV}}\xspace}
\newcommand{\TeV}{\ensuremath{\,\text{TeV}}\xspace}

\newcommand{\alphas}{\ensuremath{\alpha_\text{s}}\xspace}

\newcommand{\GF}{\ensuremath{G_\mu}}

\newcommand{\ptsub}[1]{\ensuremath{p_{\text{T},#1}}\xspace}

\newcommand{\mur}{\mu_\mathrm{R}}
\newcommand{\muf}{\mu_\mathrm{F}}

\newcommand{\pythia}{{\sc Pythia~8}\xspace}%
\newcommand{\herwig}{{\sc Herwig~7}\xspace}%
\newcommand{\sherpa}{{\sc Sherpa~3}\xspace}%

\newcommand{\newc}{\newcommand}
\newc{\bi}{\begin{itemize}}
\newc{\ei}{\end{itemize}}
\newc{\benu}{\begin{enumerate}}
\newc{\eenu}{\end{enumerate}}
\newc{\bc}{\begin{center}}
\newc{\ec}{\end{center}}
\newc{\bfig}{\begin{figure}}
\newc{\efig}{\end{figure}}
\newc{\qbar}{\bar{q}}
\newc{\go}{\tilde{g}}
\newc{\PB}{\textsc{Powheg-Box}\xspace}
\newc{\PBR}{\textsc{Powheg-Box-Res}\xspace}

\newcommand{\Recola}{{\sc Recola}\xspace}
\newcommand{\Sherpa}{{\sc Sherpa}\xspace}

\newcommand{\madgraph}{{\sc\small MadGraph5\_aMC@NLO}\xspace}
\newcommand{\rT}{{\mathrm{T}}}
\newcolumntype{.}{D{.}{.}{-1}}
\newcolumntype{d}[1]{D{.}{.}{#1}}

\colorlet{tableoverheadcolor}{gray!37.5}
\colorlet{tableheadcolor}{gray!25}
\colorlet{tablerowcolor}{gray!12.5}

\newlength{\width}
\newlength{\height}

\def\draftdate{\relax}
\def\mda{\relax}
\def\mua{\relax}
\def\mla{\relax}
\def\draft{
\def\thtystars{******************************}
\def\sixtystars{\thtystars\thtystars}
\typeout{}
\typeout{\sixtystars**}
\typeout{* Draft mode!
         For final version remove \protect\draft\space in source file *}
\typeout{\sixtystars**}
\typeout{}
\def\draftdate{\today}
\def\mua{\marginpar[\boldmath\hfil$\uparrow$]%
                   {\boldmath$\uparrow$\hfil}\color{black}%
                    \typeout{marginpar: $\uparrow$}\ignorespaces}
\def\mda{\color{red}\marginpar[\boldmath\hfil$\downarrow$]%
                   {\boldmath$\downarrow$\hfil}%
                    \typeout{marginpar: $\downarrow$}\ignorespaces}
\def\mla{\marginpar[\boldmath\hfil$\rightarrow$]%
                   {\boldmath$\leftarrow $\hfil}%
                    \typeout{marginpar: $\leftrightarrow$}\ignorespaces}
\def\Mua{\marginpar[\boldmath\hfil$\Uparrow$]%
                   {\boldmath$\Uparrow$\hfil}\color{black}%
                    \typeout{marginpar: $\uparrow$}\ignorespaces}
\def\Mda{\color{red}\marginpar[\boldmath\hfil$\Downarrow$]%
                   {\boldmath$\Downarrow$\hfil}%
                    \typeout{marginpar: $\downarrow$}\ignorespaces}
\def\Mla{\marginpar[\boldmath\hfil\textcolor{red}{$\Rightarrow$}]%
                   {\boldmath\textcolor{red}{$\Leftarrow $}\hfil}%
                    \typeout{marginpar: $\leftrightarrow$}\ignorespaces}
\overfullrule 5pt
\oddsidemargin 15mm
\marginparwidth 29mm
}

\definecolor{comment}{rgb}{0,0.3,0}
\usepackage{listings}
\newcommand{\email}[1]{\href{mailto:#1}{#1}}

\begin{document}
\begin{flushright}
LHCHWG-2025-003, CERN-TH-2025-126, FERMILAB-PUB-25-0410-T, FR-PHENO-25-08, MCNET-25-16, TIF-UNIMI-2025-14, TTK-25-19
\end{flushright}

\begin{center}{\Large \textbf{
      Higgs production via vector-boson fusion at the LHC
\\ }}\end{center}

\begin{center}
Gaetano Barone\textsuperscript{1},
Jiayi Chen\textsuperscript{2},
Stephane Cooperstein\textsuperscript{3},
Nikita Dolganov\textsuperscript{2},\\
Silvia Ferrario Ravasio\textsuperscript{4},
Yacine Haddad\textsuperscript{5},
Stefan H{\"o}che\textsuperscript{6},
Barbara J{\"a}ger\textsuperscript{7},\\
Alexander Karlberg\textsuperscript{4},
Alexander M\"uck\textsuperscript{8},
Mathieu Pellen\textsuperscript{9},
Christian T.~Preuss\textsuperscript{10},\\
Daniel Reichelt\textsuperscript{4},
Simon Reinhardt\textsuperscript{7},
Marco Zaro\textsuperscript{11}
\end{center}

\begin{center}
  {\small
{\bf 1} Physics Department, Brown University, Providence RI, USA \\
{\bf 2} Department of Physics, Simon Fraser University, Burnaby BC, Canada \\
{\bf 3} University of California San Diego, 9500 Gilman Dr, La Jolla, CA 92093, USA \\
{\bf 4} CERN, Theoretical Physics Department, 1211, Geneva 23, Switzerland\\
{\bf 5} Northeastern University, Boston, MA, USA \\
{\bf 6} Fermi National Accelerator Laboratory, Batavia, IL, 60510, USA \\
{\bf 7} Institute for Theoretical Physics, University of T\"ubingen, Auf der Morgenstelle 14, 72076 T\"ubingen, Germany\\
{\bf 8} Institute for Theoretical Particle Physics and Cosmology, RWTH Aachen University, 52056 Aachen, Germany \\
{\bf 9} Albert-Ludwigs-Universit\"at Freiburg, Physikalisches Institut, Hermann-Herder-Str.\ 3, 79104 Freiburg, Germany \\
{\bf 10} Institut f\"ur Theoretische Physik, Georg-August-Universit\"at G\"ottingen, 37077 G\"ottingen, Germany \\
{\bf 11} TIFLab, Universit\`a degli Studi di Milano \& INFN, Sezione di Milano, Via Celoria 16, 20133 Milano, Italy \\

  {\small \sf \email{gaetano.barone@cern.ch}, \email{jiayi.chen@cern.ch}, \email{stephane.b.cooperstein@cern.ch}, \email{nikita\_dolganov@sfu.ca}, \email{silvia.ferrario.ravasio@cern.ch}, \email{yacine.haddad@cern.ch}, 
  \email{shoeche@fnal.gov}, \email{jaeger@itp.uni-tuebingen.de},
  \email{alexander.karlberg@cern.ch}, \email{mueck@physik.rwth-aachen.de},
  \email{mathieu.pellen@physik.uni-freiburg.de}, \email{christian.preuss@uni-goettingen.de},
  \email{d.reichelt@cern.ch}, \email{simon.reinhardt@uni-tuebingen.de},
  \email{marco.zaro@mi.infn.it}
  }}
\end{center}

\section*{Abstract}
{\bf  
  In this article, we summarise the recent experimental measurements and theoretical work on Higgs boson production via vector-boson fusion at the LHC.
  Along with this, we provide state-of-the-art predictions at fixed order as well as with parton-shower corrections within the Standard Model at $13.6 \TeV$.
  The results are presented in the form of multi-differential
  distributions as well as in the Simplified Template Cross Section
  bins.  All materials and outputs of this study are available on
  public repositories.  Finally, following findings in the literature,
  recommendations are made to estimate theoretical uncertainties
  related to parton-shower corrections.\\ 
  }
\noindent\rule{\textwidth}{1pt}

\newpage

\tableofcontents\thispagestyle{fancy}
\noindent\rule{\textwidth}{1pt}


\newpage

\section{Introduction}

The unprecedented statistical accuracy and reduced systematics
expected during the high-luminosity phase of the Large Hadron Collider (LHC)
will make the study of the Higgs boson and the electroweak (EW) sector
of the Standard Model an endeavor in precision physics.
Higgs production via vector-boson fusion (VBF) is the mechanism with the
second highest cross section after the gluon-fusion process and
deserves particular attention due to its importance for the determination
of the Higgs boson couplings. Its
unique topology, characterised by the associated production of two
jets with large invariant mass and large rapidity separation, makes it
a relatively clean channel to study experimentally.

In order to study this mechanism in detail, it is crucial to have
dedicated precise theoretical predictions within the Standard
Model. In particular, given that VBF is typically measured in rather
exclusive phase-space regions, inclusive predictions are not
appropriate. To that end, we compile in the present study
state-of-the-art predictions for the LHC at a centre-of-mass energy of
$13.6\TeV$ at fixed order in common set-ups at the differential
level.\footnote{This is in contrast to typical predictions provided by
the LHC Higgs Working Group (LHCHWG) which typically focus on
inclusive
predictions~\cite{LHCHiggsCrossSectionWorkingGroup:2016ypw,Karlberg:2024zxx}.}
It includes next-to-next-to-leading-order (NNLO) QCD corrections at the order $\mathcal{O}(\alphas^2 \alpha^3)$,
next-to-leading-order (NLO) EW corrections at the order $\mathcal{O}(\alpha^4)$, and irreducible
contributions arising at orders $\mathcal{O}(\alphas^2 \alpha^2)$, $\mathcal{O}(\alphas^3 \alpha^2)$, and $\mathcal{O}(\alphas^4 \alpha)$.  As a by-product of our study, a comparison at
leading-order (LO) and NLO QCD accuracy between calculations performed
within and without the so-called VBF-approximation is also provided.
In addition, several representative predictions are compared for
predictions at NLO matched to parton showers (PS).  This allows us
to make recommendations for PS uncertainties for experimental
analyses following existing findings in the
literature~\cite{Jager:2020hkz,Hoche:2021mkv,Buckley:2021gfw}.
It is worth emphasising that all the results (at fixed order and with
parton showers) are obtained for multi-differential distributions as
well as for Simplified Template Cross Section (STXS)
bins~\cite{Berger:2019wnu}.

Finally, we provide a review of theoretical progress and an explicit
table with key references to be cited for specific types of
calculations.  In addition, recent experimental results by the ATLAS
and CMS collaborations are also reviewed.

In order to ensure transparency we have made all the results and
auxiliary data files publicly available at:
\begin{center}
\url{https://gitlab.cern.ch/LHCHIGGSXS/LHCHXSWG1/VBFStudyYR5} 
\end{center}
In particular, input cards, possible customisations for the computer
codes, {\sc Rivet}~\cite{Bierlich:2019rhm} routines, and histograms
are provided there.
In addition, to allow experiments to perform benchmarks of event generators we have also stored there a subset of events used in the study.

The article is organised as follows:
In Sec.~\ref{sec:review}, relevant theoretical work as well as recent experimental measurements are briefly reviewed.
The computer programs used for the present study are described in Sec.~\ref{sec:computerPrograms}.
Section~\ref{sec:setup} summarises numerical input parameters, phase space definitions, as well as the observables considered in the subsequent numerical analyses.
In Sec.~\ref{sec:FO}, fixed-order results are discussed.
In Sec.~\ref{sec:PS}, several parton-shower predictions are compared and recommendations for the assessment of parton-shower uncertainties for VBF processes are given.
Section~\ref{sec:summary} contains a summary of the study as well as some concluding remarks.

This work was carried out in the context of the LHC Higgs Working Group (LHCHWG) and in particular the VBF subgroup of the working group 1 (Higgs cross-section and branching ratios).\footnote{At \url{https://twiki.cern.ch/twiki/bin/view/LHCPhysics/LHCHWG}, a summary description of the various activities of the working group can be found.}

\section{Review of the state of the art}
\label{sec:review}
In this section, we review the theoretical and experimental status of VBF concentrating on the LHC.
In particular, by providing Tables~\ref{tab:exp-ref} and \ref{tab:th}, we highlight the corresponding references.

\subsection{Process definition}

\begin{figure}
\centering
\includegraphics[width=0.4\linewidth]{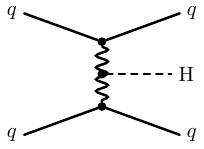}\hspace*{1cm}
\includegraphics[width=0.4\linewidth]{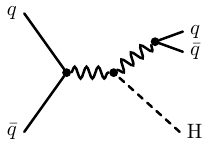}
\caption{Examples of tree-level diagrams contributing to $\process$: VBF (left) and VH (right) topologies.}
\label{diag:LO}
\end{figure}

The EW production of an on-shell Higgs boson along with two jets that we are interested in can be described as the reaction
\begin{align}
 \process .
\end{align}
At leading order (LO) in perturbation theory, the process is defined at order $\mathcal{O}\!\left(\alpha^3\right)$.
Interestingly, at this coupling order, the process contains both $t$- (and $u$-) and $s$-channel contributions.
The former constitute the \emph{VBF production} contributions.
On the other hand, the latter is usually referred to as $V\PH$ production or \emph{Higgsstrahlung}. Here, $V$ denotes either a $\PW$ or $\PZ$ boson that decays into a quark-antiquark pair.
Sample Feynman diagrams of the two production mechanisms are displayed in Fig.~\ref{diag:LO}.

An approximation, which is often used for VBF calculations, is to consider all $t$- and $u$-channel diagrams and square the contributions of these two topologies separately but do not take into account interferences between $t$- and $u$-channel diagrams. 
Also, $s$-channel contributions, as well as all interferences with them, are neglected.
These types of interference contributions are typically small~\cite{Melia:2010bm,Denner:2012dz}.
The approximation is usually referred to as the \emph{VBF approximation}, and this terminology will be used in the rest of the article.
On the other hand, what we refer to as \emph{full} computations enclose all possible contributions to the electroweak $\processfss$ production process at a given order in perturbation theory, including interferences.
We note that the predictions associated with both types of calculations are close to each other in phase space regions away from the $s$-channel resonances, \emph{e.g.}\ when  large invariant masses and rapidity differences between the two tagging jets are required. 
These are the phase space regions where VBF is measured experimentally.

\subsection{Recent experimental measurements}
\label{sec:expsummary}

Ten years after observing the Higgs boson, the current experimental accuracy reaches levels of $\approx 30\%$ on the total and inclusive cross-section in individual final states~\cite{ATLAS:2022vkf,CMS:2022dwd}. Table~\ref{tab:exp-refIncMu} summarizes the latest sensitivity on the inclusive VBF signal of both ATLAS and CMS VBF results using the Run-2 dataset at $\sqrt{s}=13 \TeV$. 
 \begin{table*}
      \footnotesize
        \begin{tabular}{l | c| c}
 & \multicolumn{2}{c}{$\sigma/\sigma_{\text{SM}}$}\\
\hline
 Channel & CMS & ATLAS \\
   \hline 
$H \to \tau\tau$ & $0.86 \pm 0.13~(\text{stat}) \pm 0.05~(\text{theo}) \pm 0.08~(\text{exp})$ \cite{CMS:2022kdi}     &  $0.93 \pm 0.12~(\text{stat}) \pm 0.11~(\text{syst})$ \cite{ATLAS:2024wfv}\\
$H \to WW^*$ & $0.71 \pm 0.26$                                                                \cite{CMS:2022uhn}    & $0.93 \pm 0.13~(\text{stat}) \pm 0.16~(\text{syst})$   \cite{ATLAS:2022ooq}\\ 
$H \to \gamma\gamma$ & $1.04 \pm 0.30~(\text{stat}) \pm 0.06~(\text{theo}) \pm 0.10~(\text{exp})$ \cite{CMS:2021kom}& $1.20 \pm 0.18~(\text{stat}) \pm 0.19~(\text{syst})$    \cite{ATLAS:2022tnm}\\ 
$H \to bb$ &  $1.01 \pm 0.50$                                                                 \cite{CMS:2023tfj}&  $0.99 \pm 0.35$                                              \cite{ATLAS:2020bhl}\\
$H \to 4l$ & $0.48 \pm 0.41~(\text{stat}) \pm 0.12~(\text{syst})$                           \cite{CMS:2021ugl}& $1.21 \pm 0.44~(\text{stat}) \pm 0.06~(\text{theo}) \pm 0.10~(\text{exp})$ \cite{ATLAS:2020rej}\\ 
 \end{tabular}
        \caption{\label{tab:exp-refIncMu} Summary of experimental sensitivity on inclusive VBF signal at the LHC from ATLAS and CMS experiments. }
        \end{table*}

This unprecedented precision achieved with the proton-proton collision data recorded at the LHC Run 1 and Run 2 allowed for a more precise investigation of these production mechanisms and VBF in particular. The STXS framework~\cite{Berger:2019wnu} provides a tool for combining the sensitivity of different Higgs boson decay channels for probing the couplings to fundamental particles at the production level. Both ATLAS and CMS have completed the experimental reach at Run-2, reaching the $10\%$ level of precision individually~\cite{ATLAS:2022vkf,CMS:2022dwd} on the combination from all available decay channels of Stage-0 STXS cross-section. 
%

The further splitting of the production phase-space into $m_{\text{jj}}$ and $p_{\text{T}}^{\text{H}}$ bins allows for setting stringent limits on new phenomena in a plethora of models and frameworks, such as Effective Field Theories, Supersymmetry, and generic coupling modifier frameworks. Beyond the channel-combined sensitivity, the LHC experiments have measured the VBF cross-section as a function of kinematic observables in single-channel decays. Fiducial measurements provide detector-independent cross-sections as a function of kinematic observables within a defined fiducial phase space.  Organized by Higgs decay channels and measurement types - STXS or fiducial cross-section measurements, Table~\ref{tab:exp-ref} summarises the references of the most recent results at Run-2 from the ATLAS and CMS experiments. 
    \begin{table*}
        \footnotesize
        \begin{tabularx}{\textwidth}{X|X|X|X|X}
            Channel  &  \multicolumn{2}{c|}{ATLAS} & \multicolumn{2}{c}{CMS} \\
            \hline
                          &  STXS/coupling & Fiducial \& differential & STXS/coupling  & Fiducial \& differential\\
            \hline
            \hline
            Combination       &   \cite{ATLAS:2022vkf,ATLAS:2024lyh}  & \cite{ATLAS:2022qef} &\cite{CMS:2022dwd} &   \cite{CMS:2018gwt} \\ 
            $\PH\to\gamma\gamma$   &  \cite{ATLAS:2022tnm}  & \cite{ATLAS:2022fnp} &  \cite{CMS:2021kom, CMS:2022wpo}& \cite{CMS:2018gwt,CMS:2022wpo} \\ 
            $\PH\to ZZ$            &  \cite{ATLAS:2020rej}  & \cite{ATLAS:2020wny} &  \cite{CMS:2023gjz,CMS:2021ugl,CMS:2021nnc}& \cite{CMS:2018gwt} \\ 
            $\PH\to WW$            &  \cite{ATLAS:2022ooq}  & \cite{ATLAS:2023pwa} &  \cite{CMS:2022uhn,CMS:2020dvg}& \cite{CMS:2020dvg}\\
            $\PH\to b\bar{b}$      &  \cite{ATLAS:2020bhl}  & & \cite{CMS:2023tfj,CMS:2024ddc,CMS:2024srp}& \cite{CMS:2024ddc}\\
            $\PH\to \tau\tau$      &  \cite{ATLAS:2024wfv}  & \cite{ATLAS:2024wfv}  & \cite{CMS:2022kdi} & \cite{CMS:2018gwt,CMS:2024jbe}\\
            $\PH\to \mu\mu$        &  \cite{ATLAS:2020fzp}  &  & \cite{CMS:2020xwi}&\\
            VBF$+\gamma$ $\PH\to b\bar{b}$ & \cite{ATLAS:2020cvh}& & &
            
        \end{tabularx}
            \caption{\label{tab:exp-ref} Summary of most recent and most precise probes of VBF production modes at the LHC. The column on the left lists the references to Simplified Template Cross Sections and couplings results, while the column on the right the fiducial and differential cross section measurements.}
    \end{table*}

Projecting the total and inclusive cross-section measurements up to the first $\text{ab}^{-1}$ of delivered luminosity at the LHC~\cite{ATL-PHYS-PUB-2022-018, CMS-PAS-FTR-18-011,Cepeda:2019klc}, their accuracy becomes limited by modeling uncertainties. The main limiting factor for the inclusive and total cross-sections remains the missing higher fixed-order calculations in both QCD and EW accuracies. However, the picture becomes more complex when going differentially and in the fiducial detector volume. The need for more segmented phase spaces stems from both a drive to measure the kinematic properties in the whole available phase space and from an experimental drive, where reconstruction techniques bring increased sensitivities in terms of background rejection when differentiating in terms of the kinetic properties of the final-state particles. The limiting factor from the modeling uncertainties stems from parton-shower modeling (including also soft physics such a multi-parton interaction and hadronisation) and matching schemes.  This is becoming a sizable contribution already for the first hundreds of $\text{fb}^{-1}$ of delivered luminosity, as for example Ref.~\cite{ATLAS:2022ooq} gives relative VBF modeling uncertainties of $12\%$, which is large compared to the total uncertainty of $18\%$ but also larger than what has been found in the present work or in Ref.~\cite{Buckley:2021gfw}.\footnote{We note that the uncertainty estimate of Ref.~\cite{ATLAS:2022ooq} also includes hadronisation effects which is not the case for the present work or Ref.~\cite{Buckley:2021gfw}.} 
Thus, looking ahead at the first $\text{ab}^{-1}$ of proton-proton collisions at the LHC, it becomes evident that a coherent effort is needed to define a clear path for understanding and reducing these uncertainties. 

\subsection{Higher-order corrections}
\label{sec:HO}

\begin{figure}
\centering
\includegraphics[width=1\linewidth]{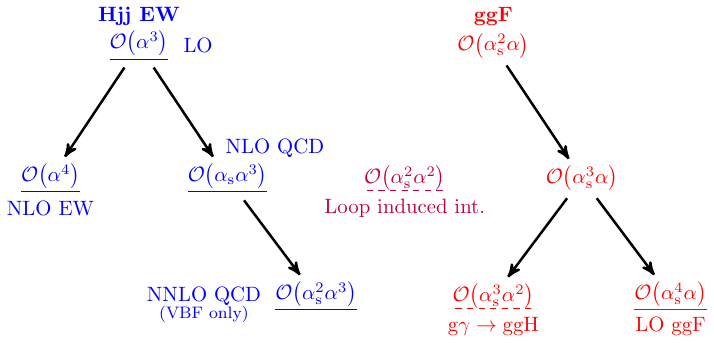}
\caption{Graphical representation of higher-order contributions computed in the present work.
See text for description.}
\label{diag:HO}
\end{figure}

In this section, we discuss several processes and orders in perturbation theory that are contributing to the production of a Higgs boson in association with two jets.
These contributions are summarised in Fig.~\ref{diag:HO}.
There, the blue contributions are the ones associated with the EW production of Hjj.
The red ones are those associated with the gluon-gluon fusion process which is considered a background.
The purple contribution cannot be unambiguously attributed to either of the two classes of process mentioned above.
The (dashed-)underlined contributions are the ones that have been (partially) calculated for the present work.

\subsubsection{QCD corrections}

\paragraph{Factorisable QCD corrections for VBF}
At NLO, within the VBF approximation, QCD corrections to the upper and lower quark lines factorise. 
This inspired the so-called ``structure-function approach''~\cite{Han:1992hr} which, by relying on the structure functions of 
deep-inelastic scattering (DIS), made it possible to compute NLO inclusive predictions for this process. Later on,
the approach was extended to the fully differential case at NLO~QCD~\cite{Figy:2003nv,Berger:2004pca} and implemented in the parton-level Monte-Carlo generators {\tt VBFNLO}~\cite{Arnold:2008rz, Baglio:2011juf, Baglio:2014uba} and {\tt MCFM}, 
and to NNLO QCD~\cite{Bolzoni:2010xr,Bolzoni:2011cu}, where gluon exchange between the two quark lines is non-zero, but still color-suppressed.
Finally, NNNLO QCD corrections were also computed for the inclusive case~\cite{Dreyer:2016oyx}.
Nowadays, two calculations are available for the differential computation of cross section at NNLO QCD accuracy~\cite{Cacciari:2015jma,Cruz-Martinez:2018rod} in  the VBF approximation.

Note that in Ref.~\cite{Asteriadis:2024nbg}, NNLO QCD corrections to both  production of a Higgs boson via VBF and its decay into a bottom-antibottom quark pair were considered for the first time.
The corrections to the full process with Higgs decay products are larger than in the inclusive case with a stable Higgs due to the constrained phase space on the decay products.
In Ref.~\cite{Behring:2025msh}, the description of the decay part was improved from NNLO QCD to NNLO QCD+parton shower accuracy.
The inclusion of all-orders resummation provided by the parton shower in the decay reduces the fiducial cross section by about $10\%$.
\begin{figure}
\centering
\includegraphics[width=0.4\linewidth]{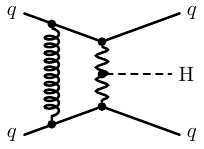}
\includegraphics[width=0.4\linewidth]{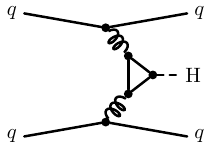}
\caption{Examples of non-factorisable corrections (left) and loop-induced contributions (right).}
\label{diag:loop}
\end{figure}

\paragraph{Non-factorisable QCD corrections for VBF}
At NNLO-QCD, the factorised picture discussed above breaks down due to
gluon exchange between the two quark lines, cf.\ Fig.~\ref{diag:loop}
(left). The one-loop diagram shown there enters already at NLO-QCD,
but at this order it enters only through the interference with the
Born diagram, and since the exchanged EW bosons do not carry colour
this contribution vanishes due to colour conservation. At NNLO-QCD the
squared diagram and also the interference between the two-loop diagram
and the Born diagram enter. Formally, they are colour-suppressed since
the exchanged gluons have to form a colour-singlet, but at the
precision at which other contributions to VBF are currently known, they
cannot be neglected.

The full computation of the non-factorisable corrections at NNLO-QCD
is not within the scope of current technology, but significant
progress has been made in computing them in the so-called eikonal
approximation. In Ref.~\cite{Liu:2019tuy} the leading eikonal term was
for the first time computed, and the corrections to VBF within typical
cuts were found to range between a few per mille and
a per cent. More differential studies were carried out in
Ref.~\cite{Dreyer:2020urf} where the calculation was also extended to
di-Higgs VBF production. Since then the original work has been
complemented with sub-leading eikonal effects and full real-emission
matrix elements~\cite{Asteriadis:2023nyl,Long:2023mvc,Bronnum-Hansen:2023vzh}.
In this study we only consider the leading eikonal contributions.

\paragraph{QCD corrections to the full EW $\process{}$ production}

As previously discussed, in the VBF approximation, one
disregards $s$-channel contributions stemming from associated VH production where the vector boson V decays hadronically,
as well as interferences between $t$- and $u$-channel contributions.
This approximation is accurate in typical VBF fiducial regions while it is not very accurate in inclusive phase space regions~\cite{Ciccolini:2007jr,Campanario:2013fsa,Bolzoni:2011cu}.
The full calculations at NLO QCD accuracy are available for up to $\PH + 3\Pj$~\cite{Campanario:2013fsa,Campanario:2018ppz}.
The limiting factor for computing the complete VBF process at NNLO QCD is the availability of the two-loop virtual corrections.

\subsubsection{EW corrections}
\label{sec:EW-corrections}

The virtual EW corrections consist of all possible one-loop insertions of EW particles to the Born process. 
On the other hand, for the real emission corrections, only photon radiation is taken into account. 
Contributions due to the emission of massive weak bosons are not included, as they are not needed to cancel infrared divergences in the virtual corrections, and can be removed from experimental analyses via appropriate vetoes.

Electroweak corrections have initially been computed for the full EW $\process$ production in Ref.~\cite{Ciccolini:2007jr}, and for the VBF approximation in Ref.~\cite{Ciccolini:2007ec}.
Typically, for current hadron colliders, the corrections are at the level of a few per cent for fiducial cross sections.
On the other hand, for differential distributions, they can become large in parts of the phase space.
Usually in the high-energy limit, they become negatively large because of Sudakov logarithms while below resonances, they are positive due to real photonic emissions.

To obtain fixed-order predictions at NLO EW accuracy, dedicated codes like {\sc Hawk}~\cite{Denner:2014cla} or multi-purpose codes like \madgraph~\cite{Frederix:2018nkq} or \sherpa~\cite{Sherpa:2024mfk} are publicly available.
Electroweak corrections to both the VBF-approximated and the full process can be obtained.

\subsubsection{Irreducible contributions}
\label{sec:irrContr}

The NLO QCD corrections to the hadronic EW process $\process$ are of the order $\mathcal{O}\!\left(\alphas \alpha^3\right)$ while the NNLO QCD corrections are order $\mathcal{O}\!\left(\alphas^2 \alpha^3\right)$.
For the process at hand, loop-induced amplitudes of order $\mathcal{O}\!\left(\alphas^2 \sqrt{\alpha} \right)$ also exist, as depicted on the right-hand side of Fig.~\ref{diag:loop}.
They can be interfered with the EW production to provide $\mathcal{O}\!\left(\alphas^2 \alpha^2 \right)$ contributions or squared to raise $\mathcal{O}\!\left(\alphas^4 \alpha \right)$ contributions.
These types of corrections are separately infrared finite and are typically rather suppressed.

The corrections of order $\mathcal{O}\!\left(\alphas^2 \alpha^2 \right)$ have been computed first in Ref.~\cite{Andersen:2006ag} in the infinite top-mass limit for the $t$ and $u$ channels.
In Ref.~\cite{Andersen:2007mp,Bredenstein:2008tm}, NLO QCD corrections to the interferences i.e.\ contributions of order $\mathcal{O}\!\left(\alphas^3 \alpha^2 \right)$ have been computed and  found to be negligible for experimental measurements.
The contributions presented here therefore constitute an update with respect to Ref.~\cite{Andersen:2006ag} by providing full top-mass dependence as well as all channels (including $s$-channel contributions).

The contributions of order $\mathcal{O}\!\left(\alphas^4 \alpha \right)$ are formally part of the NNLO QCD corrections to the production of a Higgs boson via gluon-gluon fusion.
In Refs.~\cite{Chen:2021azt,Andersen:2022zte}, state-of-the-art corrections to the order $\mathcal{O}\!\left(\alphas^4 \alpha \right)$ have been presented, unfortunately in phase-space regions not typical for VBF measurements.
In the present work, we provide the $\mathcal{O}\!\left(\alphas^4 \alpha \right)$ contribution for the quark-induced channels with full top-mass dependence which is a truly irreducible IR-finite contribution.
In addition, we also provide the complete set of contributions that include external gluons.
We refer to the latter as \emph{LO ggF} as they correspond to the LO contribution to the gluon-gluon fusion process when requiring two identified jets.

In Ref.~\cite{Harlander:2008xn}, the gluon-gluon induced contribution for Higgs production via vector-boson fusion in a loop-induced process and contributing at $\mathcal{O}\!\left(\alphas^2 \alpha^3 \right)$ was computed.
It is formally part of the NNLO QCD correction to VBF with a gluon-gluon initial state instead of the quark-quark one.
Due to its smallness, this contribution has not been reproduced here.

Finally, for completeness the loop-induced contribution $\Pg\gamma \to \Pg\Pg\PH$ appearing at order $\alpha^2\alphas^3$ has also been computed.
Due to its dependence on the photon PDF and the relevant coupling factors, it is strongly suppressed, justifying why it has never been reported publicly before.

\subsubsection{Parton showers}
Parton showers are the key component of general purpose Monte Carlo (GPMC)
event generators, such as \herwig{}~\cite{Bahr:2008pv,Bewick:2023tfi},
\pythia{}~\cite{Sjostrand:2006za,Bierlich:2022pfr}, and
\sherpa{}~\cite{Gleisberg:2003xi,Gleisberg:2008ta,Sherpa:2019gpd,Sherpa:2024mfk},
as they enable a full-fledged description of collider events
by including the dominant logarithmically-enhanced QCD corrections at all orders.

In order to improve the description of inclusive quantities, such as
total cross sections, and the modelling of hard QCD radiation, parton
showers are routinely combined with fixed-order calculations, via a
careful matching procedure that prevents the double counting of
contributions present both in the shower and in the fixed-order
calculation.
Higgs production via VBF has been available at NLO+PS accuracy for more
than a decade (see, \emph{e.g.}\ Refs.~\cite{Nason:2009ai,Frixione:2013mta}).
Comparative studies assessing the uncertainty of such generators from
parton shower and matching variations can be found, \emph{e.g.} in
Refs.~\cite{Jager:2020hkz,Buckley:2021gfw,Hoche:2021mkv}.
In general, considering these variations leads to a $10-15\%$ uncertainty
for inclusive distributions.
An important observation is the particular sensitivity of this process to
QCD colour-coherence effects, which highly suppress the presence of
radiation at central rapidities.
The use of parton showers which violate this property can lead to
fictitiously large uncertainties~\cite{Ballestrero:2018anz,Jager:2020hkz,Hoche:2021mkv}, and
must be avoided.
Beyond NLO matching, it is also possible to combine calculations with
several jet multiplicities (\emph{e.g.}\ 2 and 3) up to NLO via multi-jet
merging~\cite{Chen:2021phj,Hoche:2021mkv}.

Another direction of improvements involves the inclusion of $s$-channel contributions and related interferences~\cite{Campanario:2013fsa,Jager:2022acp}
alongside the ones defining the VBF topology.
Nowadays all the NLO+PS event generators that support VBF production
can also handle the full $\process$ EW production mode~\cite{Buckley:2021gfw}.
The inclusion of NLO EW corrections is also a currently ongoing endeavour.
So far they have been implemented only in Ref.~\cite{Jager:2022acp},
albeit they are not (yet) combined with NLO QCD corrections.

Recent years have seen also a renewed interest in the formal
accuracy of parton-shower algorithms.
In particular, the first shower with general next-to-leading
logarithmic~(NLL) accuracy for VBF (in the factorised approach) was
presented in Ref.~\cite{vanBeekveld:2023chs}.
Accurately matching NLL-accurate showers to NLO QCD calculations for
complex processes such as VBF is still work in progress,
and for this reason we refrain from including higher-logarithmic
parton-shower algorithms in our study.

The general purpose event generators and the NLO event
generators that we consider in this study are listed in
Secs.~\ref{sec:GPMC} and~\ref{sec:NLOgen}, respectively.
In this study, we focus only on the perturbative components of
GPMC simulations, although PS simulations are embedded in a
framework to consistently incorporate soft-physics effects, such as
hadronisation and multi-parton interactions.
A comprehensive study of these soft-physics effects in the context of the \herwig{}
event generator can be found \emph{e.g.}\ in Ref.~\cite{Bittrich:2021ztq},
while a first estimate in the context of the \pythia{} event generator was
performed in Ref.~\cite{Hoche:2021mkv}.

    \begin{table*}
        \footnotesize
        \begin{center}
        \begin{tabular}{c|c}
            Code  &  References  \\
            \hline
            {\bf Fixed order}  &    \\
            \hline
            VBF approximation and higher multiplicity &  \cite{Figy:2003nv,Berger:2004pca,Ciccolini:2007jr,Bolzoni:2011cu,Campanario:2013fsa,Campanario:2018ppz} \\
            NLO EW &  \cite{Ciccolini:2007jr,Ciccolini:2007ec,Figy:2010ct}  \\
            NNLO QCD (in VBF approx.) & \cite{Cacciari:2015jma,Cruz-Martinez:2018rod,Asteriadis:2021gpd,Asteriadis:2024nbg,Behring:2025msh}  \\
            N3LO QCD (inclusive) & \cite{Dreyer:2016oyx}  \\
            Non-factorisable corrections & \cite{Liu:2019tuy,Dreyer:2020urf,Asteriadis:2023nyl,Bronnum-Hansen:2023vzh,Long:2023mvc,Gates:2023iiv} \\
            Irreducible background & \cite{Andersen:2006ag,Andersen:2007mp,Harlander:2008xn} \\
            Background process & \cite{Greiner:2016awe,Greiner:2015jha,Andersen:2017kfc,Andersen:2018tnm,Andersen:2018kjg,Chen:2021azt,Andersen:2022zte} \\
            Boosted Higgs & \cite{Becker:2020rjp,Buckley:2021gfw}\\
            \hline
            {\bf Parton showers and event generators}  &    \\
            \hline
            NLO matching in VBF              & \cite{Nason:2009ai,Frixione:2013mta,Jager:2022acp} \\ 
            Multi-jet merging in VBF         & \cite{Hoche:2021mkv,Chen:2021phj}  \\
            NLO+PS uncertainties in VBF      & \cite{Jager:2020hkz,Buckley:2021gfw,Hoche:2021mkv} \\ 
            Soft-physics effects in VBF      & \cite{Hoche:2021mkv,Bittrich:2021ztq} \\              
            NLL showers in VBF               &  \cite{vanBeekveld:2023chs} \\                        
        \end{tabular}
        \caption{\label{tab:th} Summary of references for theory work in VBF at the LHC.}
        \end{center}
    \end{table*}

\section{Computer programs}
\label{sec:computerPrograms}

In this section we describe the computer programs used in our study.

\subsection{Fixed-order programs}
\label{sec:FO-codes}

\subsubsection{\sc Hawk}

{\sc Hawk}~\cite{Denner:2014cla} provides fully differential parton-level predictions for Higgs boson production in Higgs-strahlung~\cite{Denner:2011id} and vector-boson fusion~\cite{Ciccolini:2007jr,Ciccolini:2007ec} including the complete NLO QCD and EW corrections. For the VBF channel, $t$-channel, $u$-channel, and $s$-channel contributions and the corresponding interference contributions are all available. Photon-induced channels as part of the EW NLO corrections are also included. Interference contributions between VBF diagrams and gluon-fusion diagrams, as discussed in Section~\ref{sec:EW-corrections}, are also available. Partonic channels with bottom quarks in the initial or final state are only supported at leading order. External fermion masses are neglected. Anomalous Higgs boson--vector-boson couplings are supported starting from version 2.0. The code is publicly available from \url{https://hawk.hepforge.org}.

In the present work, {\sc Hawk} has been used to validate the full calculations performed by  {\sc MoCaNLO+Recola} at LO, NLO QCD, and NLO EW accuracy as well as the contribution of order $\mathcal{O}(\alphas^2 \alpha^2)$.
In addition, {\sc Hawk} has been used to check the LO and NLO QCD prediction of {\sc ProVBF} in the VBF approximation.
Beyond LO, the checks have been performed upon omitting the bottom-quark contributions.

\subsubsection{\sc ProVBF}

{\sc
  proVBF}~\cite{Cacciari:2015jma,Dreyer:2016oyx,Dreyer:2018qbw,Dreyer:2018rfu}
is a tool which computes NNLO- and N$^3$LO-QCD corrections to both
single and double Higgs production through VBF in the factorised
approximation. At NNLO it is fully differential in the kinematics of
the jets. It uses the projection-to-Born method to achieve this using
parametrised NNLO structure
functions~\cite{vanNeerven:1999ca,vanNeerven:2000uj} as implemented in
{\sc Hoppet}~\cite{Salam:2008qg} and a fully differential VBF
$H+3\mathrm{ jet}$ NLO calculation from the
\PB~\cite{Nason:2009ai,Alioli:2010xd,Jager:2014vna}. The leading
eikonal non-factorisable corrections, first computed in
Ref.~\cite{Liu:2019tuy}, are also available~\cite{Dreyer:2020urf}. The
code is publicly available from
\url{https://github.com/alexanderkarlberg/proVBFH}.

In the present work, {\sc ProVBF} has been used to provide the NNLO QCD and the leading
eikonal non-factorisable corrections in the VBF approximation.

\subsubsection{\sc MoCaNLO+Recola}

The combination {\sc MoCaNLO+Recola} relies on the flexible Monte Carlo program {\sc MoCaNLO} and the matrix-element generator {\sc Recola}~\cite{Actis:2012qn,Actis:2016mpe}.
With this at hand, any processes in the Standard Model can be computed at full NLO accuracy \emph{i.e.}\ with QCD, EW, or mixed QCD-EW corrections.

For the one-loop corrections, {\sc Recola} relies on the {\sc Collier} library \cite{Denner:2014gla,Denner:2016kdg} to numerically evaluate the one-loop scalar and tensor integrals.
On the other hand, infrared divergences are handled with the Catani--Seymour dipole formalism for all singularities of QCD and QED type~\cite{Catani:1996vz,Dittmaier:1999mb}. Additionally, there is the option to use the FKS subtraction scheme~\cite{Frixione:1995ms} for final-state QCD corrections~\cite{Denner:2023grl}.
The program has been successfully used to compute NLO QCD and NLO EW corrections to double-Higgs production via VBF~\cite{Dreyer:2020xaj} as well as several VBS processes~\cite{Biedermann:2016yds,Biedermann:2017bss,Denner:2019tmn,Denner:2020zit,Denner:2021hsa,Denner:2022pwc}.

In the present work, full predictions, \ie including all topologies and their interferences at LO, NLO QCD, and NLO EW accuracy have been obtained from {\sc MoCaNLO+Recola}.
In addition, the irreducible contributions of orders $\mathcal{O}(\alphas^2 \alpha^2)$, $\mathcal{O}(\alphas^3 \alpha^2)$, and $\mathcal{O}(\alphas^4 \alpha)$ have also been obtained from {\sc MoCaNLO+Recola}.

\subsection{General purpose event generators}
\label{sec:GPMC}

\subsubsection{\sc Herwig 7}

\herwig{}~\cite{Bahr:2008pv,Bellm:2019zci,Bewick:2023tfi} is a general-purpose event generator, known for its unique angular-ordered
parton shower~\cite{Gieseke:2003rz}, which generalises the algorithm
implemented in its predecessor \texttt{HERWIG}~\cite{Marchesini:1983bm}, and for its hadronisation cluster
model~\cite{Webber:1983if}.
\herwig{} also includes a transverse-momentum ordered dipole
shower~\cite{Platzer:2009jq}.

\herwig{} is the event generator that allows for the largest variety of
matching methods.
First of all, it supports the matching with events generated with
external NLO generators via the standard Les Houches
interface~\cite{Alioli:2013nda}.
Secondly, it implements a dedicated \texttt{Matchbox}~\cite{Platzer:2011bc} module,
which enables to combine NLO QCD calculations with both the angular-ordered and dipole showers using the {\sc Powheg}~\cite{Nason:2004rx} or {\sc MC@NLO}~\cite{Frixione:2002ik} method.%
\footnote{For processes such as Drell-Yan, also the {\sc KrkNLO} method is available~\cite{Jadach:2016qti}.}
In the case of the dipole shower, \texttt{Matchbox} also offers the
possibility to perform NLO multi-jet merging via the unitarised
merging scheme~\cite{Bellm:2017ktr}.
A dedicated study of multi-jet merging for VBF and the full EW $\PH\Pj\Pj$ process 
can be found in Ref.~\cite{Chen:2021phj}.

In the context of this study, we will consider the following NLO+PS predictions:
\begin{itemize}
\item angular-orderd parton showering on top of NLO events generated with
  \PB{} (see Sec.~\ref{sec:powhegbox}) or \madgraph{} (see Sec.~\ref{sec:madgraph}).
\item internal NLO matching in the MC@NLO scheme with both the
  angular-ordered and the dipole parton shower.
\end{itemize}
For the latter, we use {\sc Herwig7.3.0}, and dedicated matrix element providers. In particular, for this study we employ {\sc VBFNLO} and {\sc Hjets++}
for the simulation of VBF and the full EW $\PH\Pj\Pj$ production.
\begin{itemize}
  \item {\sc VBFNLO}~\cite{Arnold:2008rz, Baglio:2011juf, Baglio:2014uba,Baglio:2024gyp} is a flexible
        parton-level Monte-Carlo program for processes with EW bosons. Besides the Standard Model, also a variety of new-physics models including anomalous couplings of the Higgs and gauge
        bosons are accounted for. For the VBF-H process, it can compute NLO~QCD and EW corrections. The $\PW\PH$ production process and several irreducible backgrounds to VBF are also available.
 \item {\sc Hjets++}~\cite{Campanario:2013fsa} is a dedicated code for the calculation of the EW production of a Higgs boson accompanied by up to 3 jets at NLO QCD, including both VBF and Higgs-Strahlung contributions, as well as their interference.
\end{itemize}

\subsubsection{\pythia}
\label{sec:py8}
The \pythia package \cite{Sjostrand:2006za,Bierlich:2022pfr} is a
multi-purpose particle-level event generator, with a historically strong
focus on the modelling of soft physics.
A core component of \pythia is its hallmark string fragmentation
model \cite{Andersson:1983jt,Sjostrand:1984ic}, which models the
non-perturbative transition from coloured partons to colourless
hadrons.
The default \pythia shower, which we refer to as the ``simple shower'', is a $p_\mathrm{T}$-ordered parton shower based on DGLAP kernels~\cite{Sjostrand:2004ef}.
Rooted in the $p_\mathrm{T}$-ordered DGLAP evolution, this default option does not, however, correctly account for soft coherence effects, which are particularly important in processes such as VBF~\cite{Ballestrero:2018anz,Jager:2020hkz,Hoche:2021mkv}.
Its use in such processes is therefore discouraged.
A dipole-recoil option is, however, available for the simple shower. This replaces the independent evolution of the initial- and final-state legs in initial–final colour dipoles with a coherent, antenna-like evolution~\cite{Cabouat:2017rzi}.
In its present 8.3
series, \pythia offers the {\sc Vincia}~\cite{Brooks:2020upa} and
{\sc Dire}~\cite{Hoche:2015sya} parton-shower algorithms as
alternatives to its default shower implementation.
In either case, multi-parton interactions are simulated in a fully-interleaved 
sequence with the shower \cite{Sjostrand:2004ef}.

A wide range of multi-jet merging schemes is available with \pythia.
Internally, merging is performed in the CKKW-L scheme~\cite{Lonnblad:2011xx}
at leading order and in the UNLOPS scheme~\cite{Lonnblad:2012ix} at
next-to-leading order.
While NLO matching schemes are not implemented internally, \pythia can
be interfaced to either \madgraph in the MC@NLO~\cite{Frixione:2002ik} scheme or \PB
in the {\sc Powheg}~\cite{Nason:2004rx} scheme.
In the case of the former, the matching is tied to the simple shower with a
(non-default) global recoil scheme for final-state branchings to match the
MC@NLO counter terms implemented in \madgraph.
In the case of the latter, dedicated {\tt PowhegHooks} are available for
all three shower models, to account for the mismatch between the shower
ordering variable and the {\sc Powheg} evolution scale \cite{Corke:2010zj}.
A practically important setting of the {\tt PowhegHooks} is the {\tt POWHEG:pThard}
mode, which controls the selection of the ``hard'' scale of the {\sc Powheg} event
above which further radiation is vetoed. While in principle equivalent up
to the formal accuracy of the matching scheme, large parametrical uncertainties
of different choices have been observed for processes which
contain jets at the Born level \cite{Buckley:2016bhy,Hoche:2021mkv,ATLAS:2023kov}.
The default choice is  {\tt POWHEG:pThard = 0}, which corresponds to
using the scale provided in the event file (\texttt{scalup}), as the hard scale for
the shower evolution. Each time an emission is produced, its transverse momentum
$p_\mathrm{T}$ according to the {\sc Powheg} definition is computed and the
emission is vetoed if the shower $p_\mathrm{T}$ is larger than \texttt{scalup}.
Another suitable choice is to recalculate the {\sc Powheg} scale
as the minimal {\sc Powheg} $p_\mathrm{T}$ across the event
\cite{Nason:2013uba,Buckley:2016bhy,Hoche:2021mkv}, corresponding to the
setting {\tt POWHEG:pThard = 2}.

In the context of this study, the simple shower with dipole recoil
and the {\sc Vincia} antenna shower are applied. NLO matching is performed
in the {\sc Powheg} scheme, using event input from \PB.
The influence of the evolution-variable mismatch between \PB
and the parton showers is mitigated by the use of vetoed showers via
{\tt PowhegHooks}. 

\subsubsection{\sc Sherpa~3}
\sherpa~\cite{Gleisberg:2003xi,Gleisberg:2008ta,Sherpa:2019gpd,Sherpa:2024mfk}\footnote{The
predictions here are based on the Sherpa 3.0 release, including bugfixes that
will be made available with version 3.0.2, and have been
validated to give identical results to the patched \Sherpa version used in
\cite{Buckley:2021gfw}.}
is a general-purpose particle-level event generator. Its development began
during the late days of the LEP experiments and primarily targeted the LHC.
\sherpa includes the two automated matrix-element generators
{\sc Amegic}~\cite{Krauss:2001iv} and {\sc Comix}~\cite{Gleisberg:2008fv},
which are used in combination~\cite{Binoth:2010xt,Alioli:2013nda}
with various one-loop providers to compute the fixed-order input for
NLO matching and multi-jet merging. Matching is performed using the
S-MC@NLO method~\cite{Hoeche:2011fd} and merging is achieved
at leading order in the CKKW-L method~\cite{Lonnblad:2001iq}, and
at next-to-leading order in the MEPS@NLO method~\cite{Hoeche:2012yf}.
In this study, we consider both the VBF-induced and the full EW $\PH\Pj\Pj$ production
mode.
For this study, \sherpa implements the one-loop virtual corrections to VBF in the
structure function approximation~\cite{Buckley:2021gfw}, i.e. including vertex
corrections for the color connected quark dipoles only, and interfaces
to {\sc OpenLoops}~\cite{Cascioli:2011va,Buccioni:2019sur} for the
one-loop virtual corrections to the full EW $\process$ process.
For the predictions shown in this study we employ both \sherpa's
default parton shower~\cite{Schumann:2007mg}, sometimes referred to as the CS shower, and the {\sc Dire}
parton shower~\cite{Hoche:2015sya}. Infrared subtraction at fixed
order QCD is carried out using the Catani-Seymour dipole factorization
method~\cite{Catani:1996vz}, in the case of {\sc Dire} adapted to the
splitting kernels in this parton shower. The subtraction is
implemented in both {\sc Amegic}~\cite{Gleisberg:2007md} and Comix.

\subsection{NLO event generators}
\label{sec:NLOgen}
\subsubsection{\sc \madgraph}
\label{sec:madgraph}

The metacode {\sc \madgraph}~\cite{Alwall:2014hca,Frederix:2018nkq} makes it possible to simulate arbitrary processes including NLO QCD and EW corrections, and to perform the matching to PS with
the MC@NLO method~\cite{Frixione:2002ik} in the former case. The computation of NLO corrections relies on the 
FKS subtraction scheme~\cite{Frixione:1995ms,Frixione:1997np,Frederix:2009yq,Frederix:2016rdc} for what concerns the local subtraction
of IR divergences, and on {\sc MadLoop}~\cite{Hirschi:2011pa}, which in turns employs several one-loop 
reduction techniques~\cite{Passarino:1978jh,Davydychev:1991va,Denner:2005nn,Ossola:2006us,Cascioli:2011va,Mastrolia:2012bu} implemented in 
third-party tools~({\sc CutTools}~\cite{Ossola:2007ax}, {\sc IREGI}~\cite{ShaoIREGI}, {\sc Ninja}~\cite{Peraro:2014cba,Hirschi:2016mdz}, 
{\sc Collier}~\cite{Denner:2014gla,Denner:2016kdg}), which are either shipped with the main 
code, or can be installed at the user's request. The simulation of VBF-H production within {\sc \madgraph} has been documented in Refs.~\cite{Frixione:2013mta,Jager:2020hkz}.
As explained in the latter work,
the simulation of Higgs production via VBF at NLO-QCD accuracy can be performed issuing the commands:
\begin{verbatim}
    import model loop_qcd_qed_sm_Gmu
    generate p p > h j j $$ w+ w- z [QCD]
    output
\end{verbatim}
The syntax {\tt\$\$ w+ w- z} vetoes those diagrams with $\PW$ or $\PZ$ bosons in the $s$-channel. Furthermore, the default behaviour of {\sc \madgraph}
includes only those loops made entirely up of QCD-interacting particles,\footnote{If the EW-capable v3~\cite{Frederix:2018nkq} is employed, such a limitation
can be lifted. However, because of the tininess of these contributions, we opt for not including them.} whereas tree-level interferences are always considered.
In this study we also consider the full EW $\PH\Pj\Pj$ production mode, that can be simulated omitting  {\tt\$\$ w+ w- z}.

Finally, let us discuss the matching to PS. In principle both
\pythia~\cite{Sjostrand:2014zea,Bierlich:2022pfr} and
\herwig{}~\cite{Bahr:2008pv,Bellm:2019zci,Bewick:2023tfi} parton
showers can be employed.
However, as already discussed in Sec.~\ref{sec:py8}, it has been shown
that the global-recoil scheme employed in \pythia for which the MC@NLO
counterterm have been derived is unfit for VBF while other schemes,
such as the dipole recoil scheme~\cite{Cabouat:2017rzi}, should be
employed.
In the lack of MC@NLO counterterms compatible with such a scheme, we
only employ the {\sc Herwig7} (specifically, version
7.2.1~\cite{Bellm:2019zci}) angular-ordered
shower~\cite{Gieseke:2003rz} for matched predictions.

\subsubsection{\sc \PB}
\label{sec:powhegbox}
The \PB{}~\cite{Alioli:2010xd} is a general framework for the matching of
NLO calculations with parton-shower (PS) programs according to the
POWHEG prescription~\cite{Nason:2004rx,Frixione:2007vw}.
Dedicated implementations have been provided for the VBF-induced $\PH\Pj\Pj$
process at NLO-QCD matched to PS in Ref.~\cite{Nason:2009ai}, and for
the $\PH\Pj\Pj\Pj$ process in Ref.~\cite{Jager:2014vna}.
In Ref.~\cite{Jager:2022acp} the full EW $\PH\Pj\Pj$ production process,
including both VBF and Higgsstrahlung topologies, has been accounted
for with amplitudes provided by \Recola2{}~\cite{Denner:2017wsf} in
the framework of the \PBR{}~\cite{Jezo:2015aia}, a version of the
\PB{} specifically designed to handle processes with a complicated
resonance structure or competing topologies.
This latter implementation of the full EW $\PH\Pj\Pj$ process allows for a
matching of NLO-QCD and NLO-EW corrections to a PS generator.
In this study we consider the NLO+PS generator for the VBF-induced
contribution~\cite{Nason:2009ai}, as well as the one for the full EW
$\PH\Pj\Pj$ production~\cite{Jager:2022acp}.
Events are showered with the \herwig{} angular-ordered parton shower (v 7.2.3~\cite{Bellm:2019zci})
and with \pythia~(v~8.315)~\cite{Bierlich:2022pfr}.

\subsection{Other tools not used in this study}

For completeness, we here provide a short description of computer programs which are relevant for VBF studies but which have not been used for the present study.

\begin{itemize}
 \item {\sc HEJ} is a Monte Carlo event generator for hadronic scattering processes at high energies~\cite{Andersen:2009he,Andersen:2009nu,Andersen:2011hs,Andersen:2018tnm,Andersen:2019yzo,Andersen:2023kuj}.
 It provides all-order summation of the perturbative terms dominating the production of well-separated multiple jets at hadron colliders.
 These processes involve pure multijet production, gluon-fusion production of a Higgs boson with jets,    the production of a W boson with jets, two same-sign W bosons with jets or jets with a charged lepton-antilepton pair, via a virtual Z boson and/or photon.
 Several studies have been particularly focusing on Higgs production in association with two jets as defined in Sec.~\ref{sec:irrContr}~\cite{Andersen:2009he,Andersen:2009nu,Andersen:2017kfc,Andersen:2018kjg,Andersen:2018tnm,Andersen:2022zte}.
 The code is publicly available from \url{https://hej.hepforge.org}.
     
 \item {\sc WHIZARD} is an event generator, able to compute NLO EW corrections as well as NLO QCD corrections matched to parton shower for arbitrary processes~\cite{Kilian:2007gr,Moretti:2001zz}. The code is publicly available from \url{https://whizard.hepforge.org}.
\end{itemize}

\section{Computational set-up}
\label{sec:setup}

\subsection{Input parameters}
\label{sec:inputparams}
The present input parameters are the ones recommended by the Higgs cross section working group for run-III predictions.\footnote{These can be found at \url{https://twiki.cern.ch/twiki/bin/view/LHCPhysics/LHCHWG136TeVxsec}.}
The results obtained in the present work are for the LHC running at a
centre-of-mass energy of $\sqrt{s} = 13.6 \TeV$.  We use the
\texttt{PDF4LHC21\_40} parton distribution function
(PDF) set \cite{Cridge:2021qjj,PDF4LHCWorkingGroup:2022cjn} for quarks and gluons via the \textsc{Lhapdf} interface \cite{Buckley:2014ana}.
This PDF set employs  $\alphas(M^2_\PZ)= 0.118$ for the strong coupling constant.
For photon-induced contributions, the set \texttt{LUXqed17\_plus\_PDF4LHC15\_nnlo\_100}, which relies on the method of
\citere{Manohar:2016nzj} for the extraction of the photon distribution, is used.

The following masses and widths are used:
\begin{alignat}{2}
                  \Mt   &=  172.5\GeV,       & \quad \quad \quad \Mb &= 0 \GeV,  \nonumber \\
                \MZ &=  91.1876\GeV,      & \quad \quad \quad \GZ &= 2.4952\GeV,  \nonumber \\
                \MW &=  80.379\GeV,       & \GW &= 2.085\GeV,  \nonumber \\
                M_{\rm H} &=  125.0\GeV,       &  \GH   &=  0\GeV.
\end{alignat}

The EW coupling is fixed in the $G_\mu$ scheme \cite{Denner:2000bj,Dittmaier:2001ay} upon using
\begin{align}
  \alpha = \frac{\sqrt{2}}{\pi} G_\mu \MW^2 \left( 1 - \frac{\MW^2}{\MZ^2} \right)  \qquad \text{and}  
  \qquad   \GF    = 1.16638\times 10^{-5}\GeV^{-2}\;.
\end{align}
The numerical value thus obtained reads $\alpha = 0.75652103079904 \times 10^{-2}$.
The same convention (with different input values) has been used in several comparatives studies, \emph{e.g.}\ Refs.~\cite{Proceedings:2018jsb,Ballestrero:2018anz}.

Finally, following Ref.~\cite{Cacciari:2015jma}, for the renormalisation and factorisation scales, $\mur$ and $\muf$,  the default central values $\mu_0$ used in our calculation are obtained from 
\begin{equation}
\mu_0^2 = \frac{m_H}{2}\sqrt{\frac{m_H^2}{4}+p_{T,H}^2}, 
\end{equation}
where $m_H$ and $p_{T,H}$ are the mass and the transverse momentum of the Higgs boson, respectively. In the following, scale uncertainties are estimated by a seven-point variation of $\mur$ and $\muf$.\footnote{Notice that this choice of $\mu_0$ differs from the one adopted in Ref.~\cite{Buckley:2021gfw}, \emph{i.e.}\ $\mu_0 = H_T^2/4$, with
\begin{equation}
H_T = \sqrt{m_H^2+p_{T,H}^2} + \sum_{i \in {\rm partons}} p_{T,i}.
\end{equation}
We have, however, verified for both NLO QCD and NLO QCD+PS predictions that the two choices for the central renormalisation, factorisation, and shower starting scale
are compatible within the usual seven-point scale variations. 
We discuss the variation of these scales in the context of NLO+PS in Sec.~\ref{sec:PS}.
}

\subsection{Phase-space definitions}
\label{sec:cutsdef}
For this study, three different phase space volumes were investigated:
the STXS one~\cite{Berger:2019wnu}\footnote{It is worth mentioning that the 1.2 stage of the STXS is identical to the 1.1 stage~\cite{Berger:2922392}.}, and two fiducial ones.
It is worth pointing out that the fiducial volumes have been designed in collaboration with experimentalists of both the ATLAS and CMS collaborations.
These fiducial regions could be used for the combination of ATLAS and CMS results as well as for future comparisons of theoretical predictions with experimental measurements.

\paragraph{STXS set-up}

This setup follows the definition provided in Ref.~\cite{Berger:2019wnu}: Two jets reconstructed with the anti-$k_{\rT}$ algorithm~\cite{Cacciari:2008gp} with $R=0.4$ are required with transverse momenta 
\begin{align}
\label{eq:jet}
 \ptsub{\Pj} >  30\GeV
\end{align}
and no rapidity constraints.
In addition, the Higgs boson should be central,
\begin{align}
\label{eq:h-stxs}
 |\eta_{\PH}| < 2.5 .
\end{align}

\paragraph{Fiducial setup~(a)}

The fiducial definitions are less inclusive.
For the first one, again two anti-$k_{\rT}$ jets with $R=0.4$ are required, this time with
\begin{align}
\label{eq:jet-fida}
 \ptsub{\Pj} >  30\GeV \qquad \textrm{and} \qquad |\eta_{\Pj}| < 4.7 .
\end{align}
In addition, the two hardest jets in the rapidity range $|\eta_{\Pj}| < 4.7$ should fulfill
\begin{align}
\label{eq:jet2}
 m_{\Pj\Pj} >  300\GeV \qquad \textrm{and} \qquad |\Delta y_{\Pj\Pj}| > 2 .
\end{align}
There are no event selection requirements applied to the Higgs boson, in contrast to the STXS set-up.

\paragraph{Fiducial setup~(b)}

The definition of the second fiducial volume is similar to the first one, apart from the requirement on the jets' minimum transverse momenta,  which is changed from Eq.~(\ref{eq:jet-fida}) to 
\begin{align}
\label{eq:jet3}
 \ptsub{\Pj} >  20\GeV \qquad \textrm{and} \qquad |\eta_{\Pj}| < 4.7\,. 
\end{align}
The cuts of Eq.~\refeq{eq:jet2} on the two hardest jets are retained.

The \textbf{fiducial (a)} and \textbf{fiducial (b)} setups are rather similar, but are nonetheless considered here as the ATLAS and CMS collaborations sometimes use different transverse-momentum requirements in their measurements.

In the following, the two hardest jets fulfilling the requirements of Eq.~\eqref{eq:jet} for the STXS setup, of Eq.~\eqref{eq:jet-fida} for the \textbf{fiducial (a)} phase space, and of Eq.~\eqref{eq:jet3} for the \textbf{fiducial (b)} phase space, will sometimes be referred to as the \emph{tagging jets}.
When not stated explicitly, the two jets appearing in differential distribution are always the tagging jets.

\subsection{Observables}

In this work and in the repository associated to it, several characteristic LHC observables are considered. 
Typical one-dimensional distributions are provided, and in addition specific multi-dimensional bins are presented.

\paragraph{STXS set-up}

In the STXS set-up, the usual multi-dimensional bins (provided in Ref.~\cite{Berger:2019wnu}) in the invariant mass of the two hardest jets, the transverse momentum of the Higgs boson, and the transverse momentum of the combined Higgs boson with two jets are given.
All these bins are further divided into $\Delta \phi_{\Pj\Pj}$ bins for $[0; \pi/2]$ and $[\pi/2; \pi]$ making the distribution effectively four-dimensional. Here, the variable $\Delta \phi_{\Pj\Pj}$ is defined as the absolute value of the azimuthal angle between the two tagged jets. In the SM, the distribution is symmetric about $\Delta \phi_{\Pj\Pj}=0$.
In Sec.~\ref{sec:FO-STXS}, this distribution will only be shown while integrating over the transverse momentum of the combined Higgs boson with the two jets in order to ease readability.

\paragraph{Fiducial setups~(a) and (b)}

For these fiducial set-ups, several two-dimensional distributions are provided:  $m_{\Pj\Pj} \times \ptsub{\PH}$, $m_{\Pj\Pj} \times \Delta \phi_{\Pj\Pj}$, $m_{\Pj\Pj} \times \Delta y_{\Pj\Pj}$, and $\ptsub{\PH} \times \Delta y_{\Pj\Pj}$.
The ranges and the exact binning are provided in the repository.
The ranges are chosen so that these bins can be individually measured at the end of the high-luminosity LHC in typical experimental analyses.
These are discussed in Sec.~\ref{sec:FO-fiducial}.

\section{Fixed-order predictions}
\label{sec:FO}

In this section, fixed-order results are discussed for the fiducial volumes introduced above and the STXS setup which is particularly inclusive.
Various aspects are discussed, ranging from the VBF approximation, QCD and EW  corrections to irreducible contributions.

At fixed order, we define the best prediction as
\begin{align}
\label{eq:best-pred}
\sigma_{\rm best} = \sigma^{\rm Full}_{\rm NLO\; QCD} \times \left(1+\delta^{\rm VBF}_{\rm NNLO\; QCD} + \delta^{\rm NF\; VBF}_{\rm NNLO\; QCD}\right) \times \left(1+\delta^{\rm Full}_{\rm NLO\; EW,\; no\; \gamma}\right) + \Delta^{\rm Full}_\gamma ,
\end{align}
where $\sigma^{\rm Full}_{\rm NLO\; QCD}$ designates the NLO QCD cross section in the full calculation \emph{i.e.}\ also including the VH contributions and its interference with VBF.
The various relative corrections are defined as
\begin{align}
 \delta^i_k = \frac{\Delta^i_k }{\sigma^i_{\rm LO}} ,
\end{align}
where the $\Delta^i_k$ are the corrections  of type $k$ expressed in units of the cross section. The index $i$ indicates whether a specific correction refers to the full calculation, the VBF-approximated one or the non-factorisable corrections in the VBF approximation.
Note that the photon-induced corrections are not included in the relative EW corrections but singled out and denoted by $\Delta^{\rm Full}_\gamma$.
It is worth pointing out that this combination of the various corrections is a simple extension of what has been used in the past for inclusive predictions within the HXSWG~\cite{LHCHiggsCrossSectionWorkingGroup:2016ypw,Karlberg:2024zxx}.

This definition of the best prediction in Eq.~\eqref{eq:best-pred} is trivially extended from the cross section to differential distributions as well.
Note that in parts of the phase space where the VBF approximation is not reliable, only NLO QCD accuracy is achieved with this approach.
For example, for the STXS binning, for bins with $m_{\Pj\Pj}<350\GeV$, the NNLO QCD corrections in the VBF approximation are set to zero in Eq.~\eqref{eq:best-pred} reflecting the fact that NNLO QCD corrections are not known in this part of the phase space. 
An improvement of this approximation would require including NNLO QCD contributions for VH production where the W and Z bosons decay into jets.

As mentioned above, for the results presented in the following, the LO, NLO QCD, NNLO QCD and non-factorisable corrections in the VBF approximation have been obtained from {\sc proVBF}.
All the other full contributions/corrections, {\sc MoCaNLO+Recola} has been used to provide the results.
For all LO and NLO corrections, {\sc Hawk} has been used to successfully validate the results of both {\sc proVBF} and {\sc MoCaNLO+Recola}.

\subsection{STXS setup}
\label{sec:FO-STXS}

\begin{figure}
  \centering
    \includegraphics[width=0.5\textwidth]{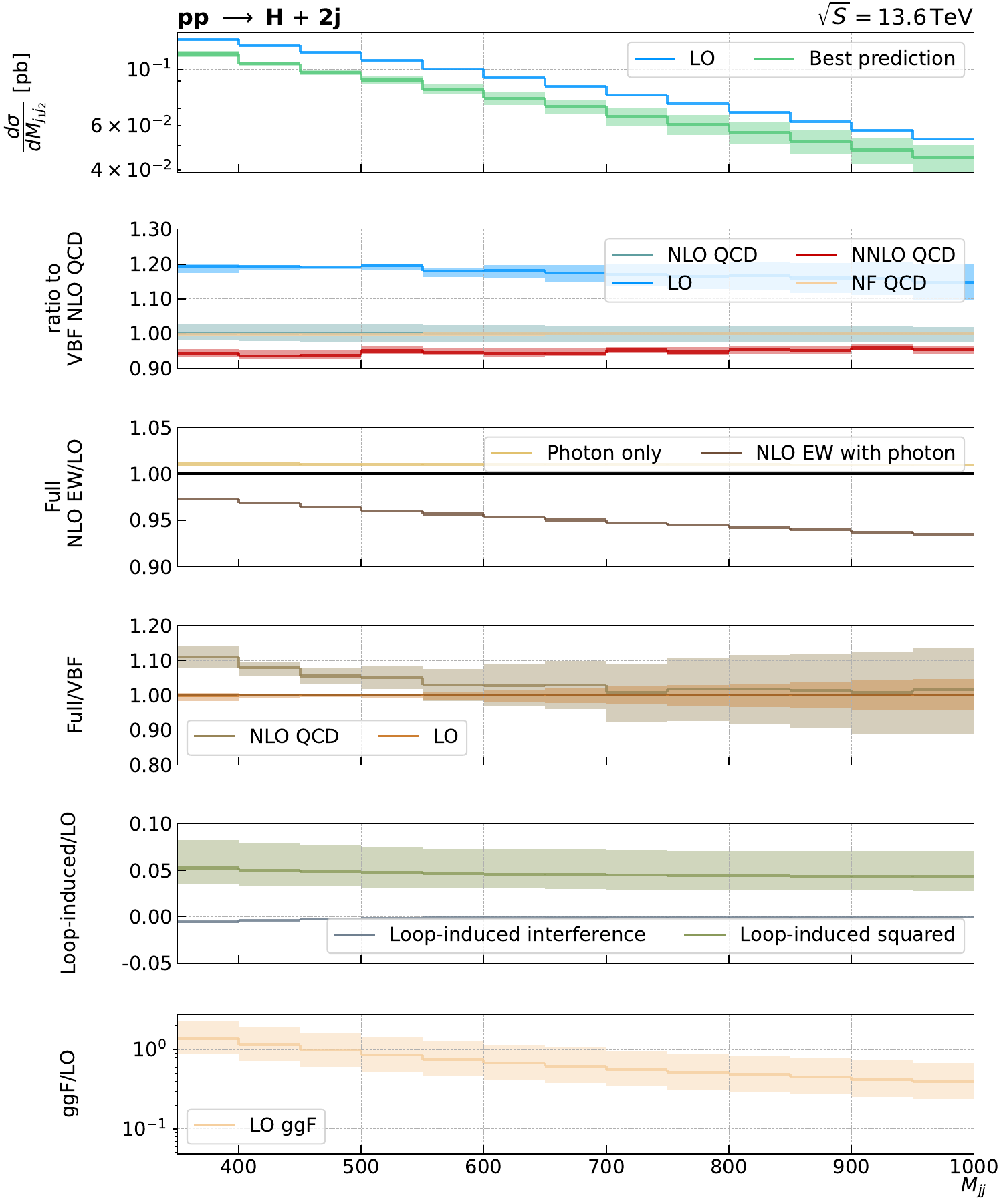}
  \caption{Differential distribution in the invariant mass of the two tagging jets for $\process$ at $13.6\TeV$ within the STXS setup Eqs.~\eqref{eq:jet}--\eqref{eq:h-stxs}. See main text for more details. 
  The bands represent the 7-point scale variations.
  \label{fig:FO-stxs-1D}
  }
\end{figure}

First, we consider the invariant-mass distribution of the tagging jets within the STXS setup of Eqs.~\eqref{eq:jet}--\eqref{eq:h-stxs}.
In Fig.~\ref{fig:FO-stxs-1D}, the uppermost panel shows the LO result within the VBF approximation as well as the best prediction obtained using the prescription of Eq.~\refeq{eq:best-pred}.
The second panel illustrates the ratio of the predictions at LO and NNLO within the VBF approximation, and the NF QCD corrections to the VBF NLO QCD result. 
The third panel of Fig.~\ref{fig:FO-stxs-1D} displays the EW corrections as well as the photon-induced contributions for the full calculation. 
The fourth panel shows the size of the non-VBF contributions at both LO and NLO QCD. 
Finally, the two lowest panels show the size of three types of contributions: loop-induced interference, loop-induced squared, and gluon-gluon fusion as described in Sec.~\ref{sec:irrContr}.
Note that in all cases, the bands represent the 7-point scale variations.

As expected, the NNLO QCD corrections are rather mild, at the level of $5\%$, while the NLO QCD ones are larger, amounting to about $20\%$ over a large range.
The non-factorisable corrections are essentially negligible and their curve is almost indistinguishable in the ratio plot.
Turning to the EW corrections, we note that, as typical for a high-energy collider such as the LHC, they grow large at high scales under the influence of Sudakov logarithms. In particular, they become as large as $-5\%$ in the invariant-mass range of about  $1\TeV$.
These corrections also include the photon-induced contribution which has been singled out and found to be at the level of $1\%$.
Interestingly, this contribution is rather constant over the entire considered phase-space region.

We now turn to a discussion of the VBF approximation.
It can be seen that at low invariant masses, the VBF approximation is very close to the full computation at LO.
On the other hand, at NLO QCD, the two computations can differ by up to $10\%$ at $\mjj=350\GeV$ while converging to within a per cent above invariant masses of $700\GeV$.
Such a behaviour is due to real QCD radiation contributions for processes with hadronically decaying heavy gauge boson, in the present case $pp\to \PH(V\to \Pj\Pj)$.
While configurations with an invariant mass of the jets around the mass of the W or Z boson are forbidden at LO due to event selection, real radiation contributions can lift these requirements, hence making $V\PH$ contributions significant.
Such a pattern has been first observed for vector-boson fusion in~\cite{Ballestrero:2018anz,Denner:2020zit} and for top-quark processes in~\cite{Denner:2017kzu,Denner:2023grl}.

Finally, the interference of the EW production with the loop-induced production is essentially negligible as it is below $1\%$ over the full range.
On the other hand, the squared loop-induced contribution is reaching about $5\%$ at $\mjj=350\GeV$ to stabilise slightly above $4\%$ above invariant masses of $600\GeV$.
We recall that these contributions are for the quark-induced channels and constitute an irreducible contribution to the EW production $\process$ that will be measured experimentally if not subtracted.

\begin{figure}
  \centering
    \includegraphics[width=0.6\textwidth]{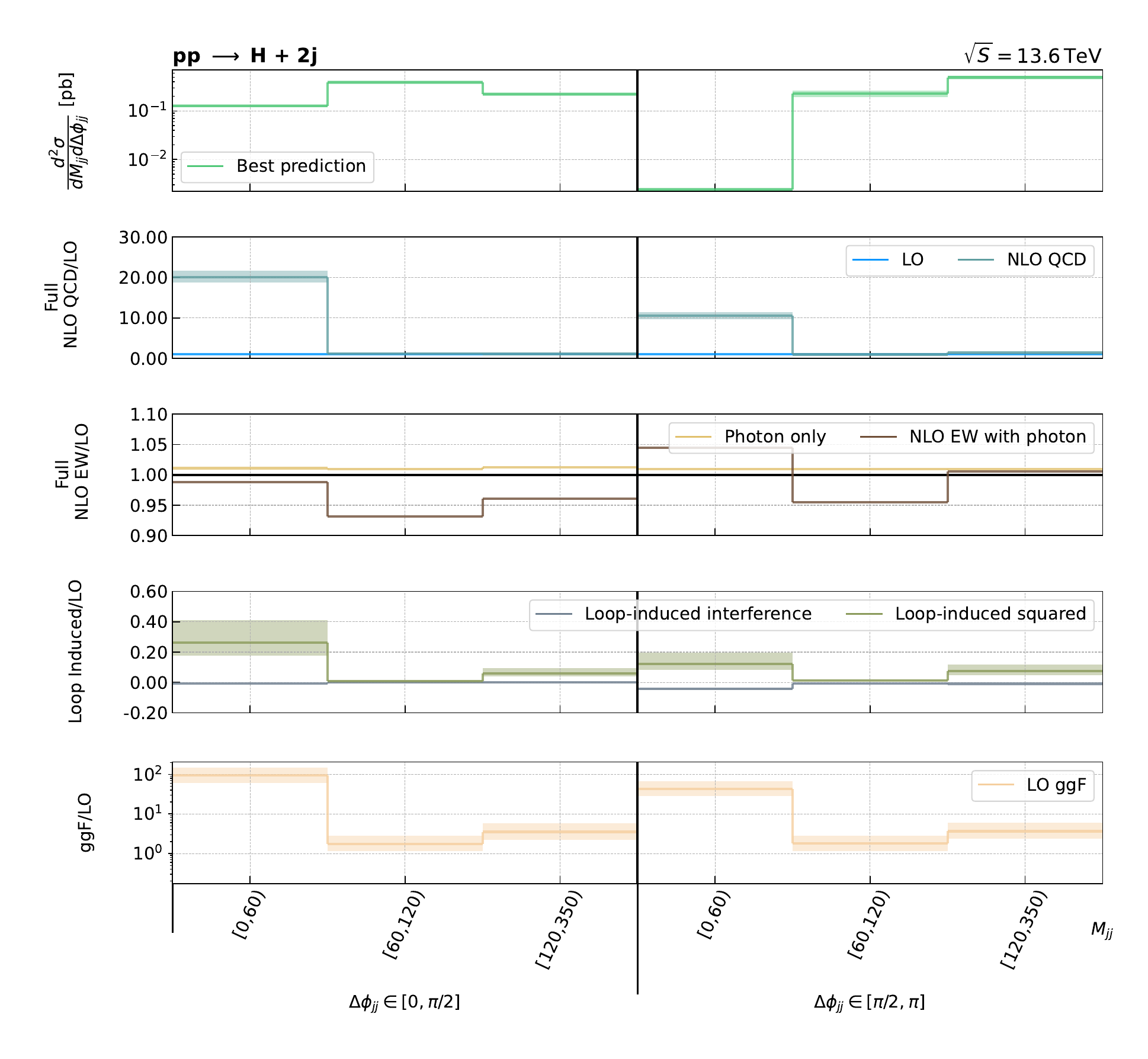}
  \caption{Multi-dimensional STXS bins for $\process$ at $13.6\TeV$. See main text for more details.  
  \label{fig:FO-stxs-bins1}
  }
\end{figure}  

\begin{figure}
  \centering
    \includegraphics[width=0.9\textwidth]{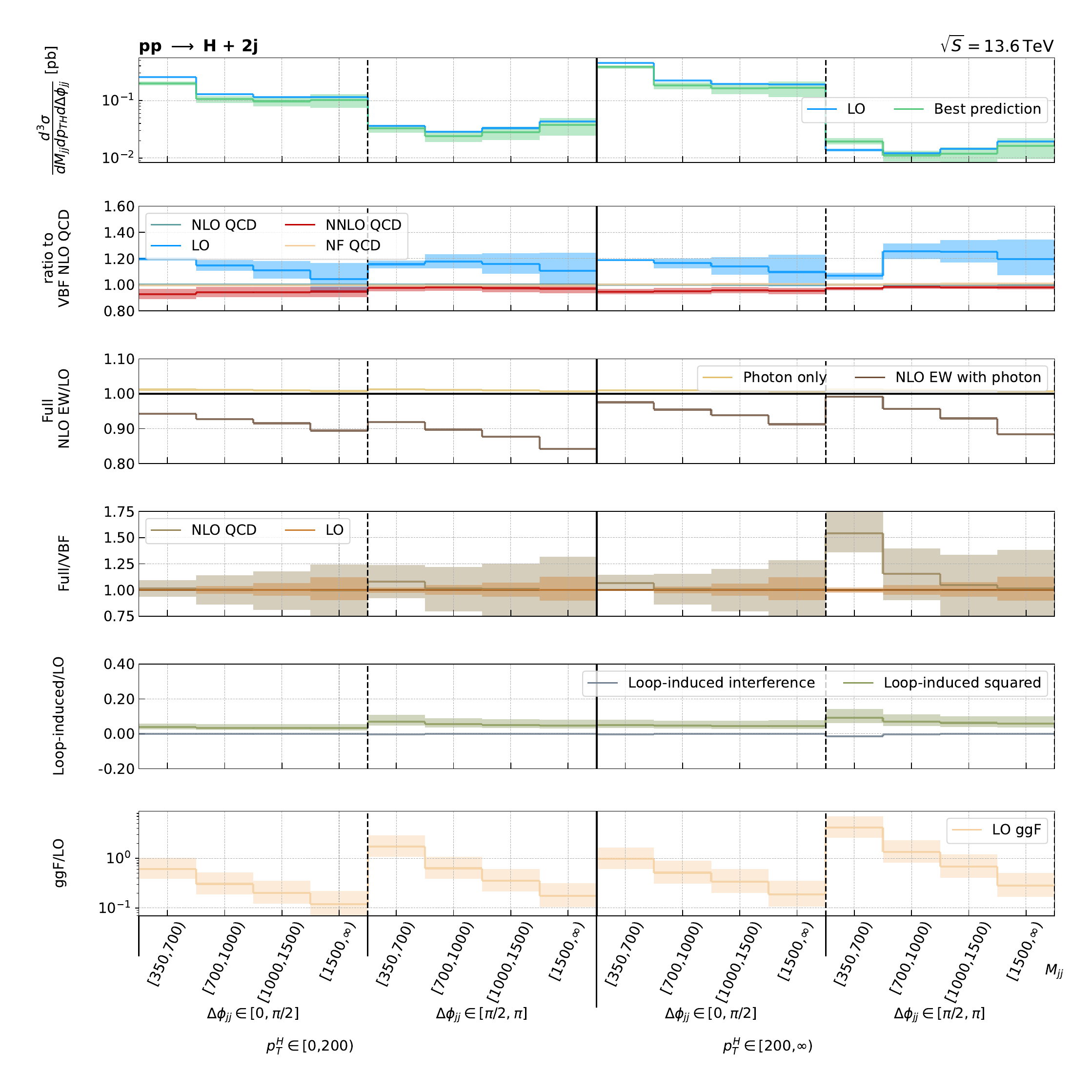}
  \caption{Multi-dimensional STXS bins for $\process$ at $13.6\TeV$. See main text for more details.
  \label{fig:FO-stxs-bins2}
  }
\end{figure}

In Figs.~\ref{fig:FO-stxs-bins1} and \ref{fig:FO-stxs-bins2}, the STXS observables of Ref.~\cite{Berger:2019wnu} are presented with panels for the individual types of corrections as in Fig.~\ref{fig:FO-stxs-1D}.
In Fig.~\ref{fig:FO-stxs-bins1}, the two-dimensional distribution in the invariant mass of the two tagging jets and their azimuthal angle difference are shown.
In particular, in Fig.~\ref{fig:FO-stxs-bins1}, the bins for the invariant mass of the two hardest jets below $350\GeV$ are displayed.
This part of the phase-space is mostly dominated by $V\PH$ production, justifying why no VBF predictions are provided.
It is interesting to observe that QCD corrections for the bins with an invariant mass  below the  masses of the W/Z~bosons (\emph{i.e.}\ $[0,60]$~GeV), the QCD corrections are gigantic, being 10~to~20 times larger than the LO predictions.
At LO, such phase-space regions cannot feature a resonant weak boson and therefore exhibit very low cross sections.
At NLO, thanks to real radiation contributions new phase space regions are opening up, allowing the weak bosons to be resonant.
The EW corrections do not show distinctive features while the photon-induced contribution remains  constant in all bins.
In this phase-space region, the loop-induced contributions are very much suppressed.
This is not the case for the gluon-gluon fusion contribution which there is larger  than the EW production by almost two orders of magnitude.
This is simply due to the gluon luminosity at low energy at the LHC.
It is worth pointing out that the STXS bins are also divided according to the transverse momentum of the Higgs boson and two jets.
We have not shown this splitting here for aesthetic reasons (the bin $[25,+\infty[$ is zero at Born level) but these bins are provided in the repository.
In addition to the original STXS prescription (stage 1.1), we have added two bins in $\Delta \phi_{\Pj\Pj}$ ($[0,\pi/2]$ and $[\pi/2,\pi]$).

In Fig.~\ref{fig:FO-stxs-bins2}, the three-dimensional distribution in the invariant mass of the two tagging jets, their azimuthal angle difference, and the transverse momentum of the Higgs boson are shown.
In particular, the STXS bins with $m_{\Pj\Pj}>350\GeV$ are shown.
The QCD corrections are moderate for all bins.
The photon-induced contributions stay at the per-cent level over the full range, while full EW corrections become  larger negative (about $-15\%$) with increasing invariant mass.
It is worth noting that the VBF approximation reproduces  the full EW production cross sections  well for these bins of high-invariant jet mass at both LO and NLO QCD, apart from the bin with  $m_{\Pj\Pj}\in[350,700]$, $p_{{\rm T}, \PH} > 200\GeV$, and $\Delta \phi_{\Pj\Pj} \in [\pi/2,\pi]$. 
There, the full EW contribution differs by $50\%$ from the VBF-approximated  one as in this bin a large $V\PH$ contribution occurs when real radiation is present.
The loop-induced contributions are also small in all considered regions.
Finally, the gluon-gluon-fusion contribution is rather large in all bins, and it culminates in the bins with high transverse momentum of the Higgs boson and $\Delta \phi_{\Pj\Pj} \in [\pi/2,\pi]$. 
Again the splitting of STXS bins according to $p_{{\rm T}, \PH\Pj\Pj}$  is not shown here,  but is available in the public repository of our results.

\subsection{Fiducial regions}
\label{sec:FO-fiducial}

In this section, fiducial cross sections as well as two-dimensional distributions are discussed.
For the plots, the various panels show the same types of contributions as in the previous section.
Given that \textbf{fiducial (a)} and \textbf{fiducial (b)} phase spaces are very similar, we here only discuss results for the former.

\begin{table*}
  \begin{center}
    \begin{tabular}{c||c|c}
     Order  & $\sigma [\pb]$ & $\delta [\%]$  \\
     \hline
     \hline
     {\bf VBF approx. } &  \\
     \hline
     LO [$\alpha^3$] & 2.479(1) & - \\
     NLO QCD [$\alpha^3\alphas$] & -0.300(3) & -12.1 \\
     NNLO QCD [$\alpha^3\alphas^2$] & -0.088(7) & -3.55 \\
     NNLO QCD non-fact.\ [$\alpha^3\alphas^2$] & -0.0053766(7) & -0.22 \\
     \hline
     {\bf Full} &  \\
     \hline
     LO [$\alpha^3$] & 2.4772(1) & - \\
     NLO QCD [$\alpha^3\alphas$] & -0.2648(10) & -10.7 \\
     NLO EW [$\alpha^4$] & -0.14759(5) & -5.95 \\
     Photon induced [$\alpha^4$] & 0.021286(2) & +0.86 \\
     Loop induced int.\ [$\alpha^2\alphas^2$] & -0.003301(1) & -0.01 \\
     Loop induced squ.\ [$\alpha\alphas^4$] & 0.136403(9) & 5.5 \\
     LO ggF [$\alpha\alphas^4$] & 1.7751(6) & 71.5 \\
     $\Pg\gamma \to \Pg\Pg\PH$ [$\alpha^2\alphas^3$] & $4.3(1)\times 10^{-6}$ & 0.0 \\
     \hline
     {\bf Best prediction} &  \\
     \hline
     $\process$ & 2.041(9) & -17.6
    \end{tabular}
  \end{center}
  \caption{
  Cross sections for the \textbf{fiducial (a)} phase space.
  The various contributions and corrections are described in detail in Sec.~\ref{sec:HO}.
  The value of the best prediction is obtained from Eq.~\eqref{eq:best-pred}.
  }
  \label{table:crossSection}
\end{table*}

\paragraph{Cross sections}
First, in Table~\ref{table:crossSection} we discuss the fiducial cross sections for  \textbf{fiducial (a)}.
In the upper part of the table, the numbers for the VBF approximation are collected.
It is interesting to observe that the NLO QCD corrections are at the level of $12\%$ while the NNLO QCD ones amount to about $25\%$ of their size, as expected from power-counting arguments.
On the other hand, the non-factorisable corrections at NNLO QCD are at the per-mille level and are therefore negligible for phenomenological considerations.
In particular, while we include them in our best predictions, they actually do not play any significant role.
In the second part of the table, \emph{full} predictions in the sense that all $t,u,s$ contributions as well as their interferences are considered.
The full LO predictions only differ insignificantly from those in the VBF approximation.
The NLO QCD corrections are also of the same order in both approaches, even if the full result is slightly lower.
The NLO EW corrections, including photon-induced contributions, are about $-5\%$ while the photon-induced contributions are below the per-cent level.
As for the non-factorisable corrections, interference contributions between the EW production and the loop-induced ones are at the per-mile level.
On the other hand, the loop-induced contributions squared, retaining only the quark channels which are therefore an irreducible background to the EW production of $\PH\Pj\Pj$, are of the order of the NLO EW corrections, but exhibit an opposite sign.
Considering also partonic channels with external gluons, the LO gluon-gluon fusion contribution with two identified jets is particularly large, at the level of $71\%$, as already observed in Ref.~\cite{Buckley:2021gfw}.\footnote{Note that in Ref.~\cite{Buckley:2021gfw}, the ggF contribution is even dominant over the EW production given that there, no invariant-mass and rapidity-difference cuts are applied to the tagging jets.}
These contributions are typically not part of the signal and subtracted in experimental analyses, which makes therefore precise theoretical predictions for ggF crucial even for VBF measurements.
Finally, the $\Pg\gamma \to \Pg\Pg\PH$ process is completely negligible due to the small photon PDF and coupling factor suppression. Compared, for instance, to the related  $\Pg\Pg \to \Pg\Pg\PH$ which is dominating the gluon-gluon fusion contribution, the $\Pg\gamma \to \Pg\Pg\PH$ cross section comes with an extra factor of $\alpha/\alphas$. 
Due to its smallness, the $\Pg\gamma \to \Pg\Pg\PH$ contribution is not discussed further and not shown in the differential distributions.

\begin{figure}
  \centering
    \includegraphics[width=\textwidth]{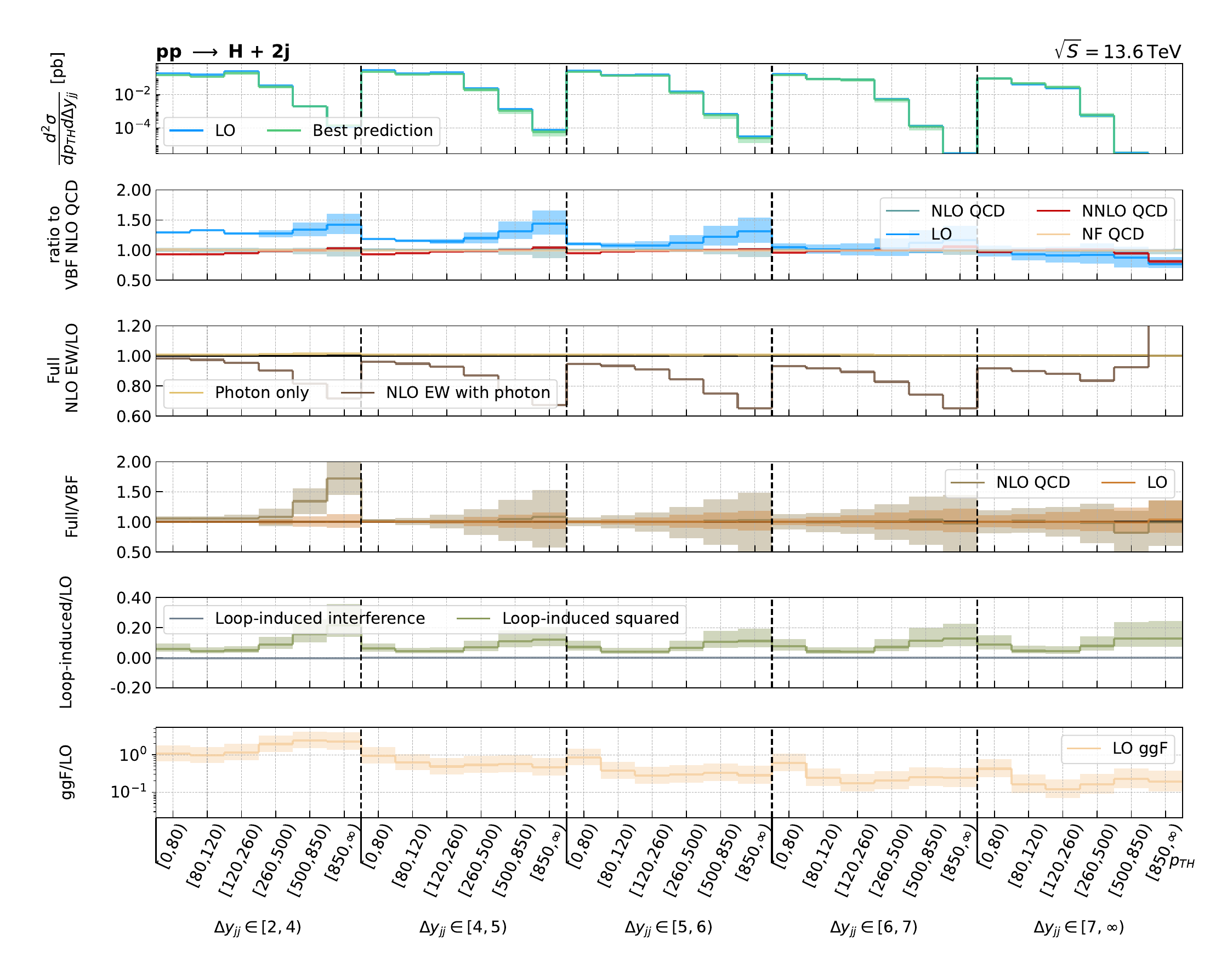}
    \caption{Differential distribution for $\process$ at $13.6\TeV$ for the \textbf{fiducial (a)} phase space for the transverse momentum of the Higgs boson and the rapidity difference of the two jets. See main text for more details.}
      \label{fig:FO-setup2a-2D-pTH-DYijj}
\end{figure}

\paragraph{Differential distributions}
In Fig.~\ref{fig:FO-setup2a-2D-pTH-DYijj},
the double differential distribution in the transverse momentum of the Higgs boson and the rapidity difference of the two jets is displayed.
Interestingly, NLO QCD corrections are largest for small rapidity differences of the two jets.
They then tend to grow larger for larger Higgs transverse momentum.
At NNLO QCD, the behaviour is opposite where the corrections tend to become smaller for larger rapidity differences.

As expected, NLO EW corrections grow negatively larger for larger transverse momentum of the Higgs boson that correspond to the high-energy limit where Sudakov logarithms become dominant.
The photon-induced contribution is almost constant over the full range at the level of $1\%$.
The two highest transverse momentum bins for $\Delta y_{\Pj\Pj} \in [7, \infty ($ exhibit a distinct behaviour.
As these bins are highly suppressed (by several orders of magnitude), their Monte Carlo errors are very large, rendering the numbers for the EW corrections unreliable there.

As explained above,  the $V\PH$ contribution is sizeable when the transverse momentum of the Higgs boson is large, \emph{i.e.}\ when the Higgs boson is recoiling against the two jets that are most likely to originate from the decay of a heavy gauge boson.
At LO the effect is not visible due to the imposed selection cuts, but it becomes pronounced at NLO QCD with an impact reaching up to $70\%$ in parts of phase space.
This effect has already been reported in Ref.~\cite{Buckley:2021gfw}.

The interference of the loop-induced  with the pure tree-level EW amplitude is negligible.
This is not the case for the loop-induced squared contribution that can be larger than $20\%$ of the full EW production of $\PH\Pj\Pj$ for large Higgs transverse momentum.
In general, the latter contributions are between $5\%$ and $10\%$ across the phase space.

\begin{figure}
  \centering
    \includegraphics[width=\textwidth]{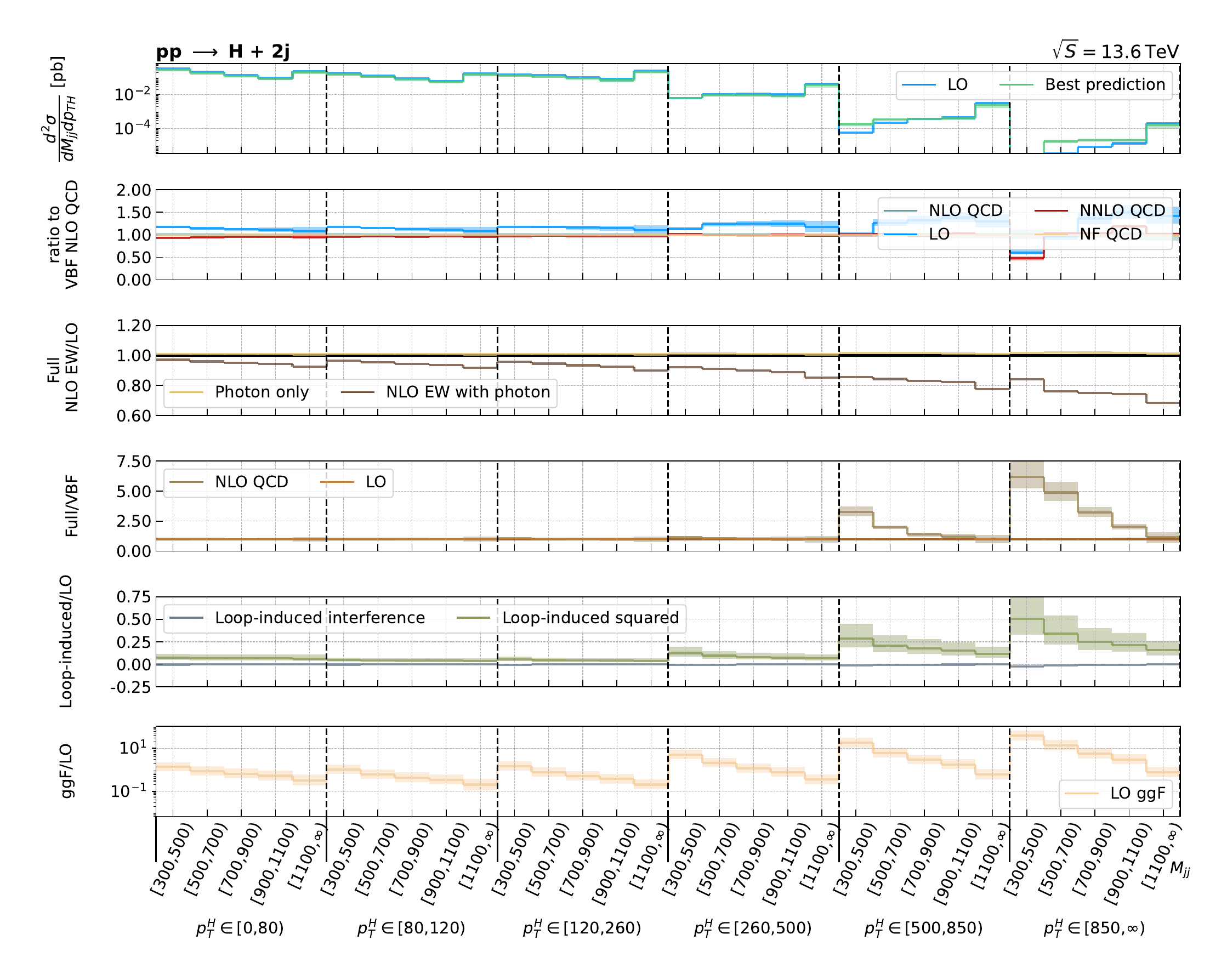}
  \caption{Differential distribution for $\process$ at $13.6\TeV$ for the \textbf{fiducial (a)} phase space for the transverse momentum of the Higgs boson and the invariant mass of the two jets. See main text for more details.
  \label{fig:FO-setup2a-2D-pTH-Mjj}
  }
\end{figure}

Let us now turn to Fig.~\ref{fig:FO-setup2a-2D-pTH-Mjj}, which shows the two-dimensional distribution  in the invariant mass and the transverse momentum of the Higgs boson.
As observed previously, the NLO QCD corrections are larger at large transverse momenta of the Higgs boson while the NNLO QCD ones become milder in these regions.

The EW corrections show similar features as in previous distributions. In particular, the quark-induced contributions are becoming large in the high-energy limit under the influence of Sudakov logarithms while the photon-induced contributions are constant throughout.

The third-to-lowest panel illustrates that the $V\PH$ contribution is particularly enhanced for large transverse momenta of the Higgs boson due to the recoil effect described above~\cite{Ballestrero:2018anz,Denner:2020zit,Denner:2017kzu,Denner:2023grl}.
The effect can be as large as almost $600\%$ in some bins.
In the very same region  the loop-induced squared contributions are also the largest. 

\paragraph{Parton-distribution functions}
Finally, given recent developments in approximate N3LO PDF sets~\cite{McGowan:2022nag,NNPDF:2024nan,MSHT:2024tdn}, we have computed LO and NLO QCD predictions using the newly introduced PDF sets.
The idea here is to quantify the numerical impact of the newly introduced PDF sets for a realistic experimental setup.
The results are tabulated in Tables~\ref{table:aN3LO_QCD_full},\ref{table:aN3LO_QCD_vbf}, and \ref{table:aN3LO_EW_full}.
At LO, we observe a difference of $3.4\%$ between the nominal \texttt{PDF4LHC21\_40} set and the approximate N3LO PDF set~\cite{MSHT:2024tdn}.
The difference between the nominal \texttt{PDF4LHC21\_40} and the NNLO QCD set is smaller and is of $0.8\%$.
For, NLO QCD in the full calculation, the relative corrections are essentially unchanged and amount to around $-10.5\%$ in both cases with differences at the per-mille level.
The same hold true for the EW corrections and photon-induced contributions as well.
For the VBF-approximated calculation, the shift at LO is at the level of $3.2\%$, in line with the full calculation.
As in the full calculation, the NLO QCD corrections in the VBF approximation using different PDF sets are virtually identical, with differences only at the per-mille level, as shown in the Table~\ref{table:aN3LO_QCD_vbf}.
Further investigations of the interplay between approximate N3LO and fixed-order calculations are left for future work.

\begin{table*}
  \begin{center}
    \begin{tabular}{c||c|c|c}
     Order & \texttt{PDF4LHC21\_40} & \texttt{MSHTxNNPDF\_NNLO\_qed} & \texttt{MSHTxNNPDF\_aN3LO\_qed} \\
     \hline
     \hline
     $\sigma^{\rm Full}_{\rm LO} [\pb]$        & 2.4772(1) & 2.4967(1) & 2.5606(1) \\
     \hline
     $\Delta^{\rm Full}_{\rm NLO\,QCD} [\pb]$  & -0.2648(10) & -0.262(1) & -0.266(1) \\
     \hline
     $\delta^{\rm Full}_{\rm NLO\,QCD} [\%]$   & -10.7 & -10.5  & -10.4 \\
    \end{tabular}
  \end{center}
  \caption{
  Predictions at LO and NLO QCD for the full calculation for the \textbf{fiducial (a)} phase space for different PDF sets.
  Note that \texttt{MSHT20xNNPDF40} is shorten to \texttt{MSHTxNNPDF} in the table.
  }
  \label{table:aN3LO_QCD_full}
\end{table*}

\begin{table*}
  \begin{center}
    \begin{tabular}{c||c|c}
     Order & \texttt{PDF4LHC21\_40} & \texttt{MSHTxNNPDF\_aN3LO\_qed} \\
     \hline
     \hline
     $\sigma^{\rm VBF}_{\rm LO} [\pb]$        & 2.479(1) & 2.5587(1) \\
     \hline
     $\Delta^{\rm VBF}_{\rm NLO\,QCD} [\pb]$  & -0.300(3) & -0.3030(3) \\
     \hline
     $\delta^{\rm VBF}_{\rm NLO\,QCD} [\%]$   & -12.1 & -11.8 \\
    \end{tabular}
  \end{center}
  \caption{
  Predictions at LO and NLO QCD for the VBF calculation for the \textbf{fiducial (a)} phase space for different PDF sets.
  Note that \texttt{MSHT20xNNPDF40} is shorten to \texttt{MSHTxNNPDF} in the table.
  }
  \label{table:aN3LO_QCD_vbf}
\end{table*}

\begin{table*}
  \begin{center}
    \begin{tabular}{c||c|c|c}
     Order & \texttt{PDF4LHC21\_40} & \texttt{MSHTxNNPDF\_NNLO\_qed} & \texttt{MSHTxNNPDF\_aN3LO\_qed} \\
     \hline
     \hline
     $\sigma^{\rm Full}_{\rm LO} [\pb]$        & 2.4772(1) & 2.4967(1) & 2.5606(1) \\
     \hline
     $\Delta^{\rm Full}_{\rm NLO\,EW} [\pb]$  & -0.14759(5) & -0.14826(3) & -0.15238(3) \\
     \hline
     $\delta^{\rm Full}_{\rm NLO\,EW} [\%]$   & -5.95 & -5.94  & -5.95 \\
     \hline
     $\Delta^{\rm Full}_\gamma [\pb]$   & 0.021286(2) & 0.021453(2) & 0.021887(2) \\
     \hline
     $\delta^{\rm Full}_\gamma [\%]$   & +0.86 & +0.86  & +0.85 \\
    \end{tabular}
  \end{center}
  \caption{
  Predictions at LO and NLO EW for the full calculation for the \textbf{fiducial (a)} phase space for different PDF sets. The photon-induced contribution is singled out.
  Note that \texttt{MSHT20xNNPDF40} is shorten to \texttt{MSHTxNNPDF} in the table.
  }
  \label{table:aN3LO_EW_full}
\end{table*}

 \section{Parton shower}
\label{sec:PS}

\subsection{Comparison between fixed order and parton shower predictions for VBF}
\label{sec:nlopscmp}

In this section, we compare the predictions obtained using the event generators described in Secs.~\ref{sec:GPMC} and~\ref{sec:NLOgen}. Specifically, we focus on the \textbf{fiducial (b)} phase space region defined in Sec.~\ref{sec:cutsdef}.  Results for the \textbf{fiducial (a)} region have similar behaviour. 
 
 \begin{figure}[t!]
 \begin{subfigure}[h!]{0.5\textwidth}
  \includegraphics[width=\textwidth]{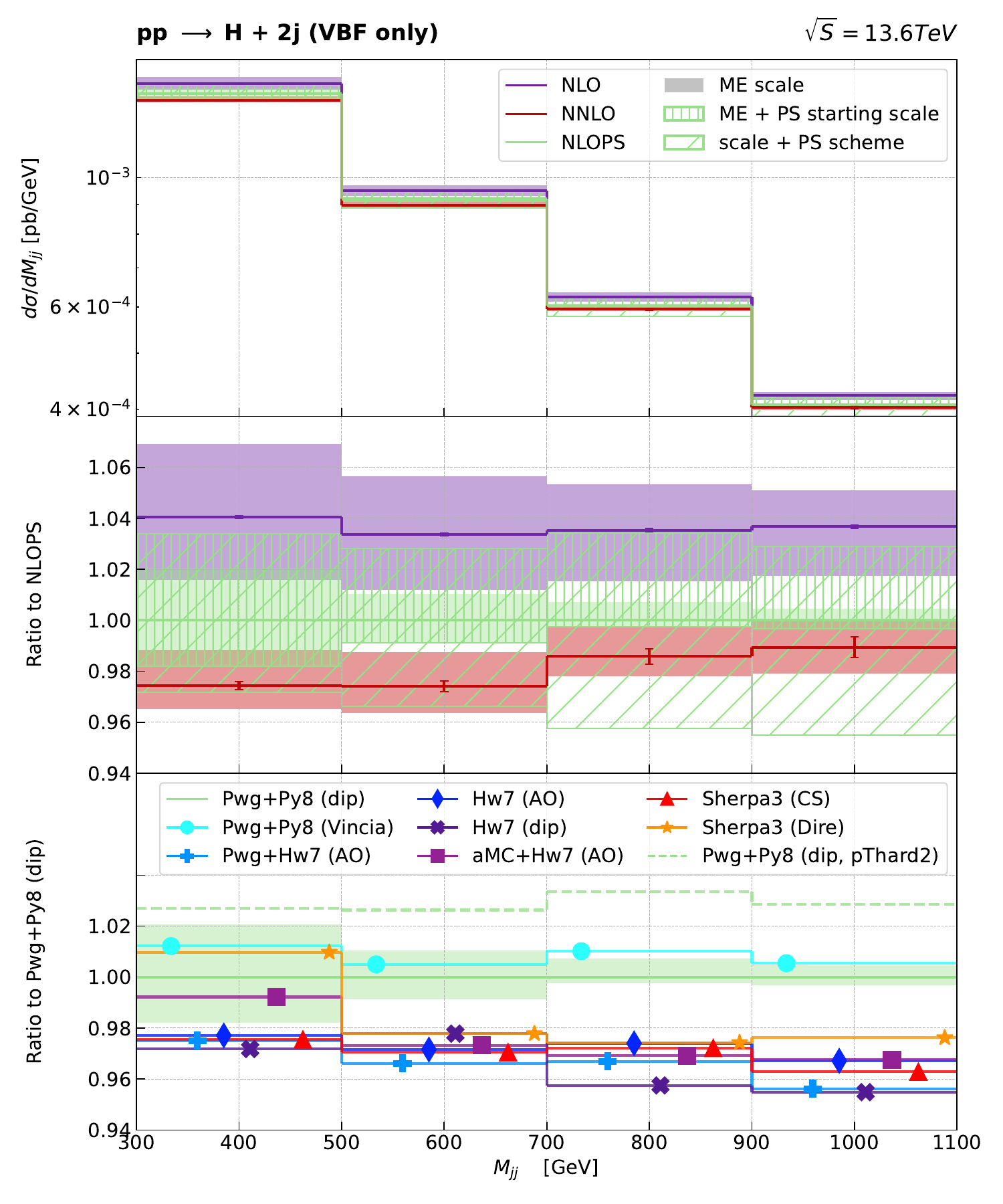}
  \caption{ $p_{T,H}<120$~GeV}
 \end{subfigure}%
  \begin{subfigure}[h!]{0.5\textwidth}
  \includegraphics[width=\textwidth]{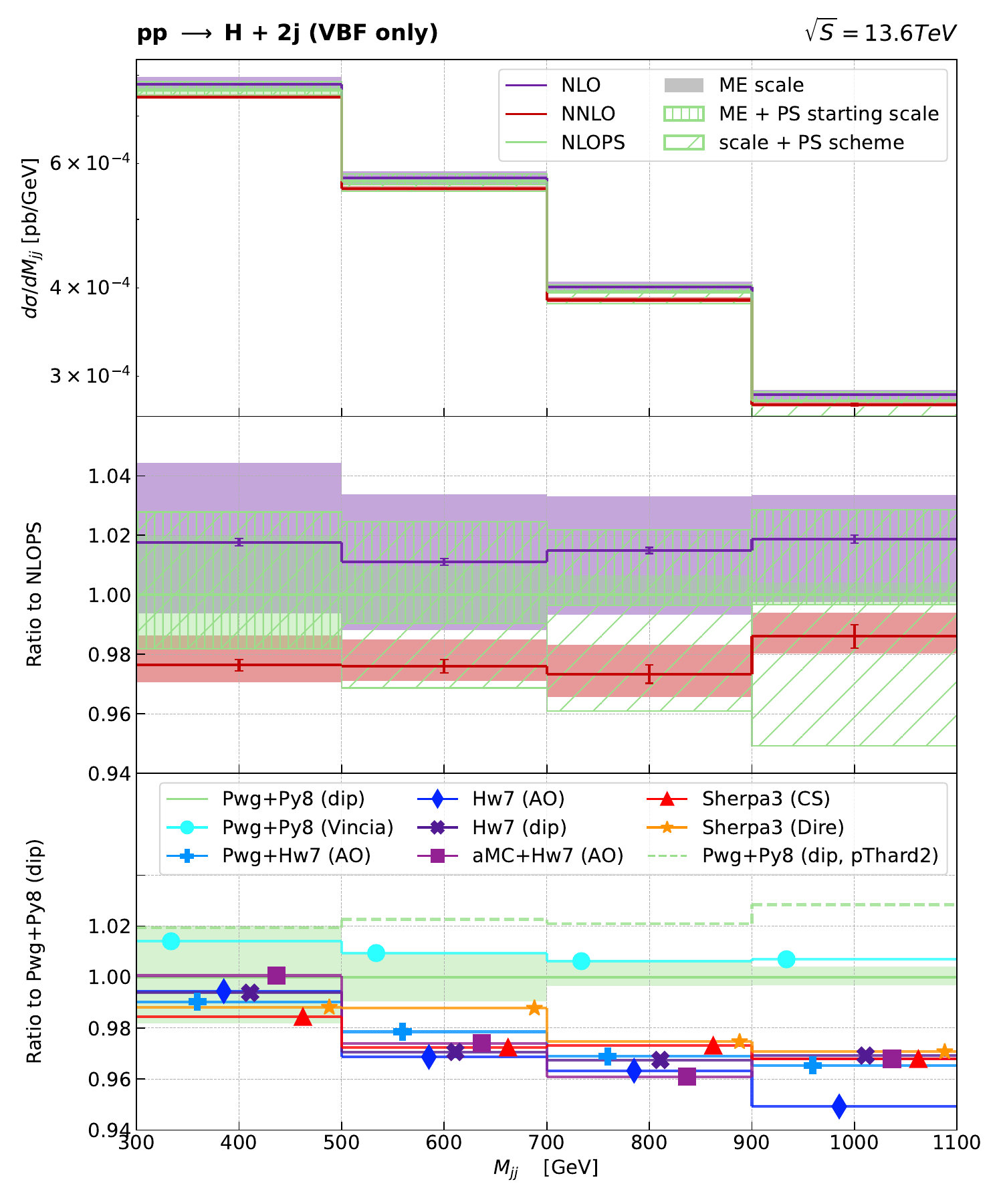}
    \caption{120~GeV $<p_{T,H}<$260~GeV}
 \end{subfigure}
 \caption{
  Fixed-order (NNLO and NLO) and NLOPS predictions within the VBF approximation for the dijet invariant mass within the \textbf{fiducial (b)} phase space, in two regions of Higgs transverse momentum.
The solid bands represent the 7-point variation of the factorisation and renormalisation scales in the hard matrix elements, while for the vertical hashed green bands additionally the shower starting scale of the reference generator is varied. 
For the total NLOPS uncertainty, shown as diagonal hashed green bands,  the scale variation and generator uncertainties are combined as detailed in our recommendations in Sec.~\ref{sec:recommendations}. 
The middle panels illustrate the ratio between the respective fixed-order and the central NLOPS predictions.
The lower panels show the spread of the predictions obtained with the various NLOPS generators considered in this study. 
}
\label{fig:MJJ-PTH}
\end{figure}

 \begin{figure}[t!]
 \begin{subfigure}[t]{0.5\textwidth}
  \includegraphics[width=\textwidth]{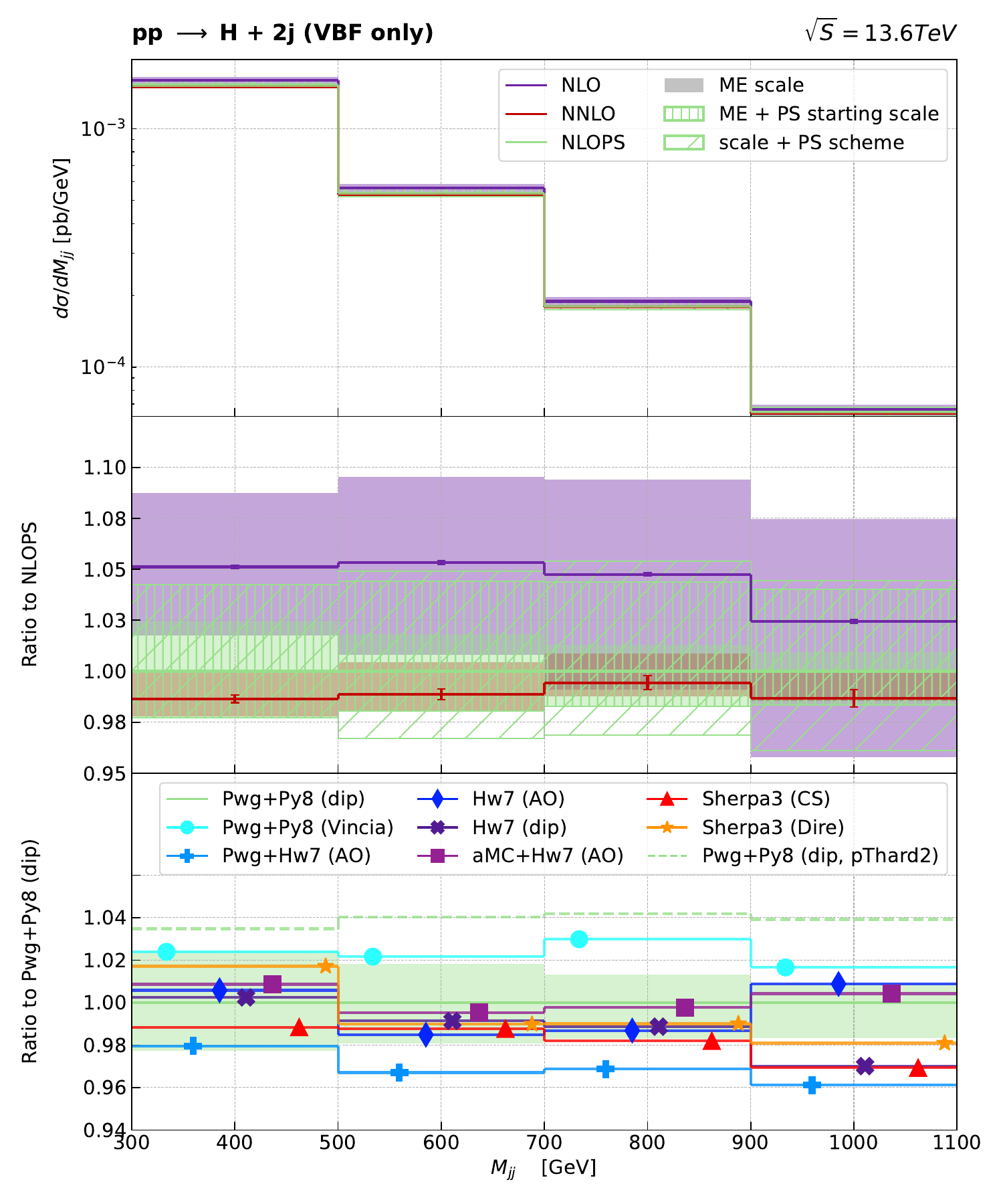}
  \caption{$2<|\Delta y_{\Pj\Pj}|<4$}
 \end{subfigure}%
  \begin{subfigure}[t]{0.5\textwidth}
  \includegraphics[width=\textwidth]{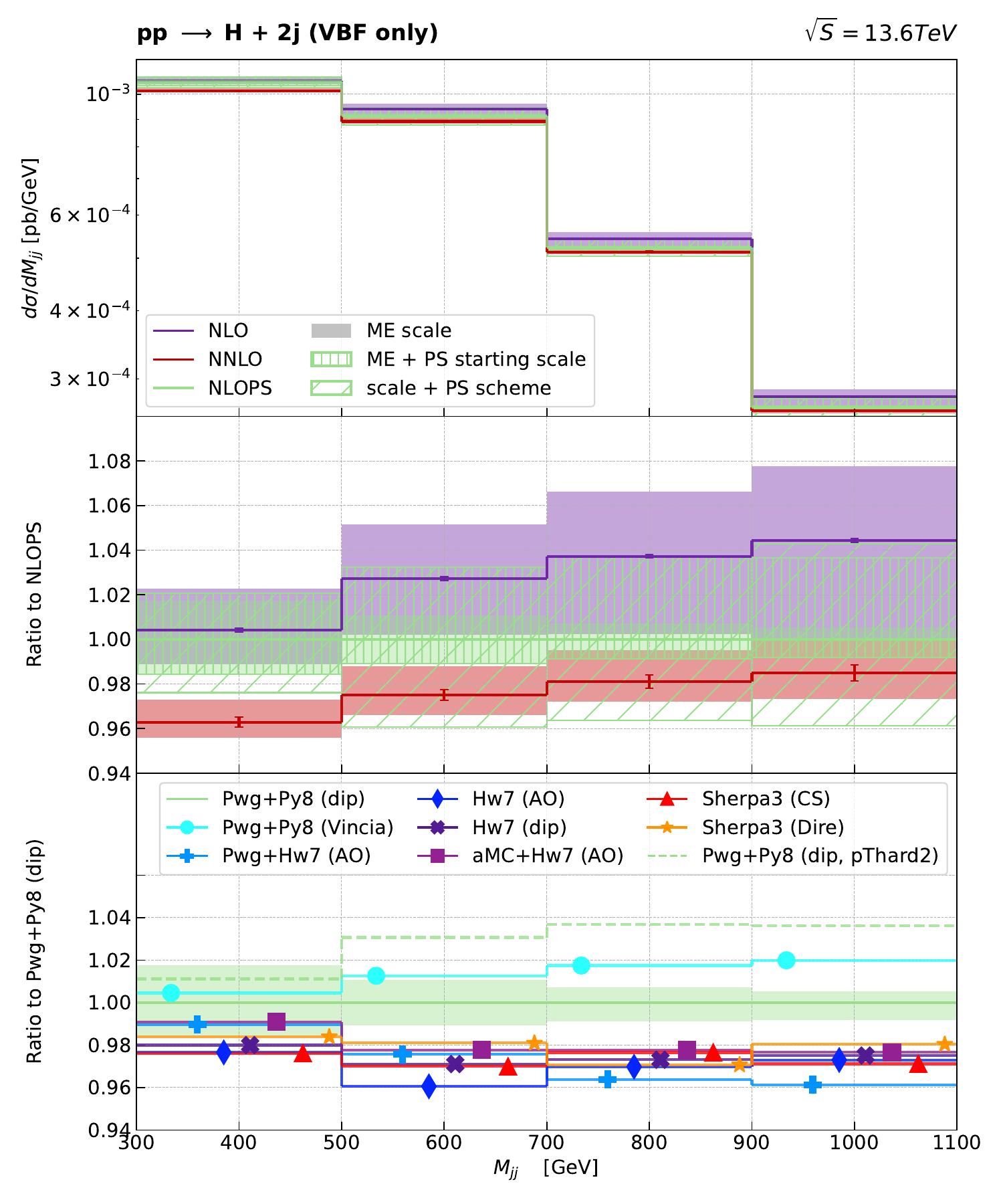}
  \caption{$4<|\Delta y_{\Pj\Pj}|<5$}
 \end{subfigure}
 \caption{As in Fig.~\ref{fig:MJJ-PTH}, but for two regions of dijet rapidity separation.} 
\label{fig:MJJ-DYJJ}
\end{figure}  

 \begin{figure}[tb!]
 \begin{subfigure}[h!]{0.5\textwidth}
  \includegraphics[width=\textwidth]{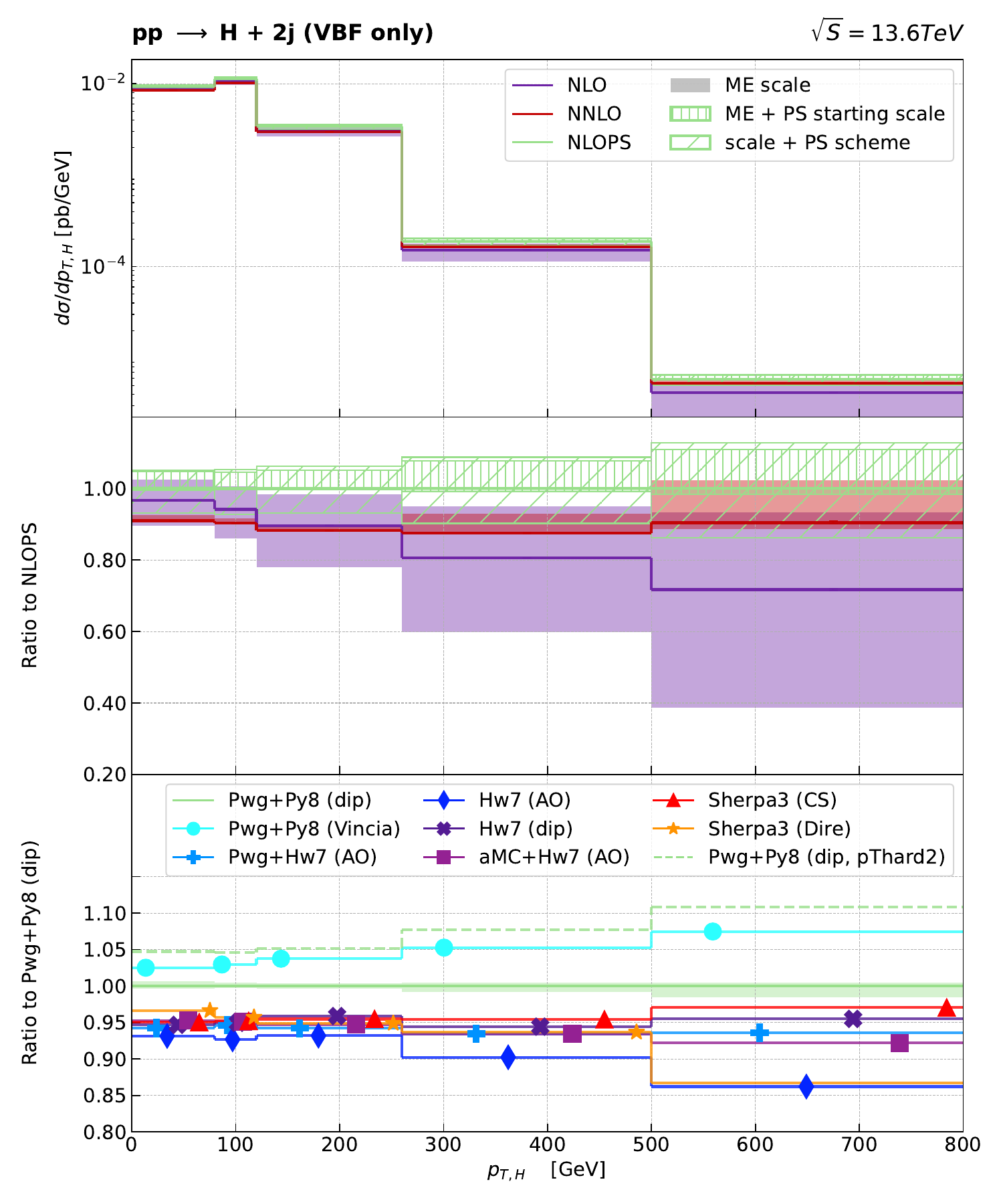}
  \caption{two reconstructed jets}
 \end{subfigure}%
  \begin{subfigure}[h!]{0.5\textwidth}
  \includegraphics[width=\textwidth]{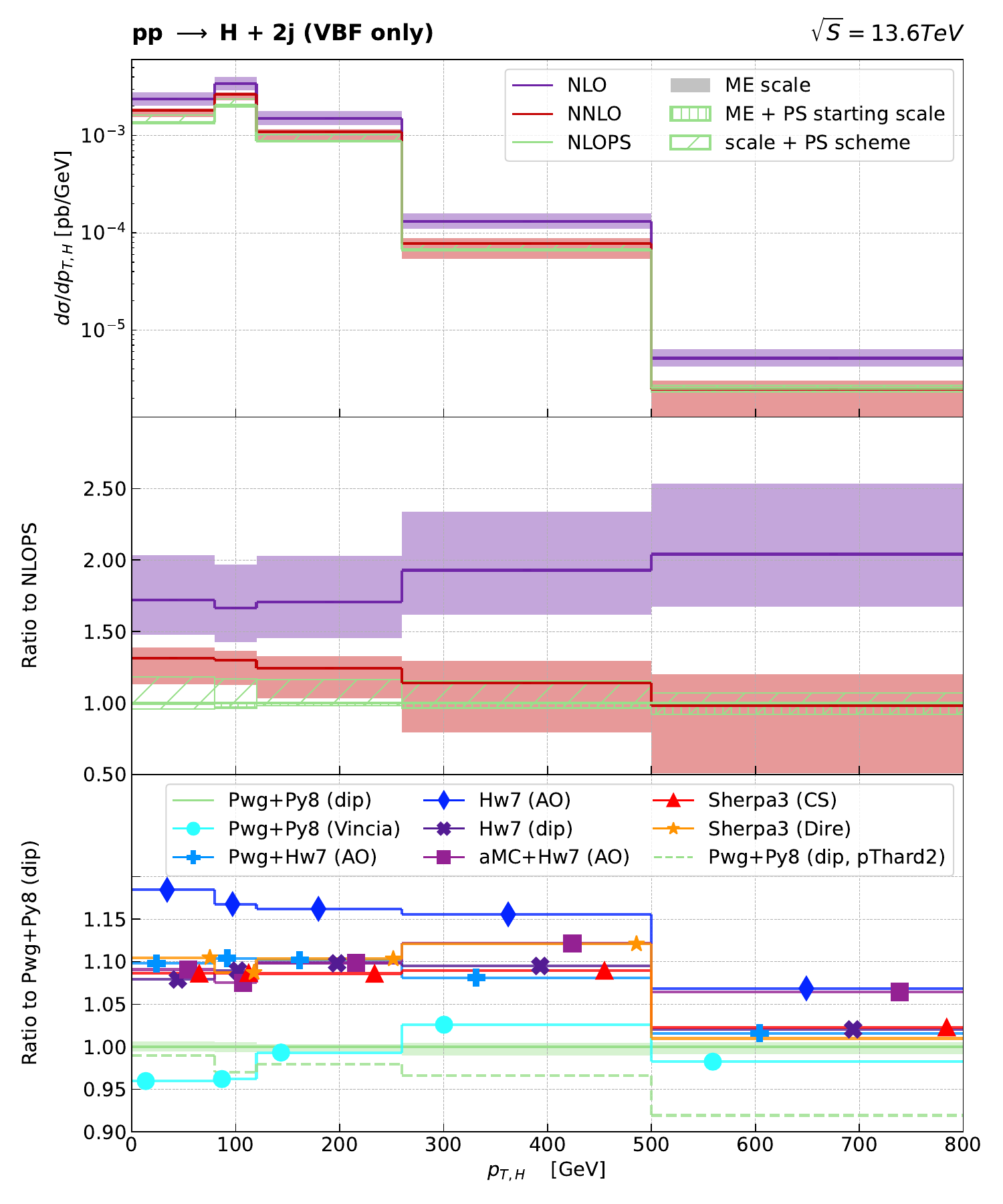}
    \caption{three reconstructed jets}
 \end{subfigure}
 \caption{
 Higgs transverse momentum within the \textbf{fiducial (b)} phase space region, requiring exactly two~(left) or three~(right) reconstructed jets.
The naming of the curves is the same as in Fig.~\ref{fig:MJJ-PTH}.
}
\label{fig:PTH-NJETS}
\end{figure}  
 
We compare NLO QCD~(violet), NNLO QCD~(red) and NLOPS~(green) predictions for both inclusive~(Figs.~\ref{fig:MJJ-PTH} and \ref{fig:MJJ-DYJJ}) and exclusive~(Fig.~\ref{fig:PTH-NJETS}) observables for illustration.
As done in the previous sections, the theoretical uncertainty associated with NLO and NNLO QCD calculations is represented by a solid band corresponding to the scale variations (dubbed \emph{ME scale} in the legend). 

At NLOPS, as our reference we use the VBF \PB{} implementation~\cite{Nason:2009ai} showered with \pythia{} using a dipole recoil and the option {\tt POWHEG:pThard = 0}.
For this prediction, we show again the scale variations in the hard matrix element (green solid band).
The total scale uncertainty associated with the reference NLOPS generator also comprises the uncertainty stemming from the choice of the hard scale in the parton shower evolution (see Sec.~\ref{sec:py8}), which we vary by using the {\tt POWHEG:pThard = 2} option.
This uncertainty is given by the hatched band labelled \emph{ME + PS starting scale}.
Finally, the total uncertainty  of the NLOPS prediction also includes the envelope of all the NLOPS generators considered in this study (scale + PS scheme).
The breakdown into several NLOPS generators, including shower starting scale for the reference prediction (dashed green line), is shown in the bottom panel.

In Fig.~\ref{fig:MJJ-PTH}, we show the dijet invariant mass $M_{\Pj\Pj}$, requiring a small  ($p_{\rm T, \PH}<80~\GeV$, left) and a moderate  ($80~\GeV<p_{\rm T, \PH}<120~\GeV$, right)  Higgs transverse momentum. 
In Fig.~\ref{fig:MJJ-DYJJ}, we instead show the dijet invariant mass requiring a moderate ($2<|\Delta y_{\Pj\Pj}|<4$, left) and a large ($4<|\Delta y_{\Pj\Pj}|<5$, right) dijet rapidity separation.
The dijet invariant mass, Higgs transverse momentum, and the dijet rapidity separation are standard inclusive observables for VBF-like processes, meaning that we expect a good description from the fixed-order predictions.
Since these observables involve jet reconstruction (directly or through the application of jet-selection cuts), the parton showering will also affect the predictions.
We find that in general the central NLOPS predictions lie between the NLO and the NNLO QCD ones.

Furthermore, the central NNLO QCD predictions lie within the total uncertainty band around the NLOPS central prediction.
The only exception is the lowest $M_{\Pj\Pj}$ bin in the presence of a boosted Higgs (right panel of Figs.~\ref{fig:MJJ-PTH}) or for very large dijet separation (right panel of Fig.~\ref{fig:MJJ-DYJJ}).
This behaviour was also observed in Ref.~\cite{Buckley:2021gfw}.

The NNLO uncertainty is of the order of $2\%$, while the NLO one is roughly twice as large.
The NLOPS uncertainty estimated by varying the matrix-element and shower starting scales is smaller than the corresponding NLO theoretical uncertainty.
However, once the variation of the NLOPS generator is included, the two uncertainties become comparable.
This indicates that only varying the parameters within a single shower algorithm may yield an underestimate of the true perturbative uncertainty.

We now consider distributions that are more sensitive to the presence of additional radiation in the final state.
In Fig.~\ref{fig:PTH-NJETS}, we illustrate the transverse momentum of the Higgs boson requiring exactly two (left panel) or three (right panel) reconstructed jets.
The first distribution can still be considered inclusive, as it is well defined at LO, whereas the second is exclusive, requiring the presence of an additional radiation jet.
Both distributions develop terms enhanced by large logarithms $L=\log(p_{\rm T,\PH}/p_{\rm T,\Pj}^{\min})$ at all-orders in $\alphas$. Thus, when $\alphas L^2 \approx 1$, \emph{i.e.}\ around $p_{\rm T,\PH}=500\GeV$, the accuracy of the fixed-order prediction is lost.
The distribution in the left panel is of particular interest given the proposal in Ref.~\cite{Buckley:2021gfw} of introducing a jet veto procedure in order to reduce the contamination from the gluon-fusion background.
In the region with $p_{T,H}\ll 500$~GeV, (N)NLO calculations 
still retain (N)NLO accuracy.
We notice that NLOPS predictions overshoot both the NLO and the NNLO QCD results, but for boosted-Higgs topologies, the NLOPS curve tends to be closer to the NNLO than the NLO one.
We also observe that requiring only 2 jets leads to a larger scale uncertainty, reaching $20\%$ in the hard tail.
In this case, the NLOPS uncertainty is much smaller than the NLO one, but much bigger than the NNLO one. 
This is somewhat expected, given the sensitivity to additional radiations.
It is also worth noting that in the last bin, sensitive to all-orders corrections that can be much bigger than the nominal NNLO prediction, the error band of the fixed-order calculation clearly does not reflect the true uncertainty.

As expected, requiring 3 jets~(right panel of Fig.~\ref{fig:PTH-NJETS}) is even more sensitive to the radiation pattern provided by the showers.
In the region with $p_{T,H}\ll 500$~GeV, (N)NLO calculations 
only retain (N)LO accuracy.
In this case, parton shower uncertainties are again of the order of 20\%, and they are of the same order as the NNLO ones for small-moderate $p_{\mathrm{T},\PH}$ while for a boosted Higgs the NNLO uncertainty substantially increases.
It is worth noting that the NLOPS prediction is in agreement with the NNLO QCD curve, but not with the NLO QCD one.

\subsection{VH interference effects}

 \begin{figure}[t!]
  \begin{subfigure}[t]{0.5\textwidth}
    \includegraphics[width=\textwidth]{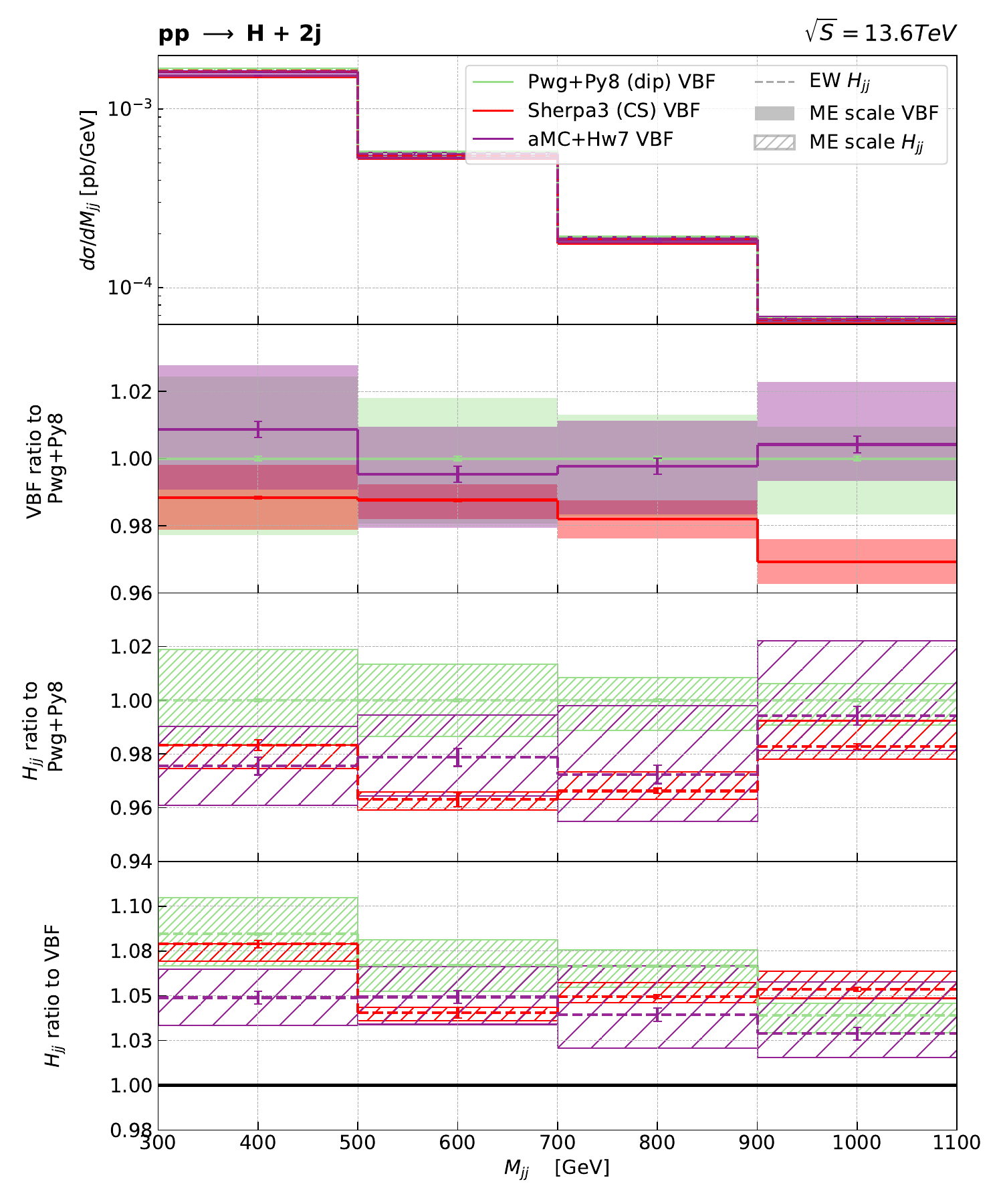}
  \caption{$2<|\Delta y_{\Pj\Pj}|<4$}
 \end{subfigure}%
  \begin{subfigure}[t]{0.5\textwidth}
 \includegraphics[width=\textwidth]{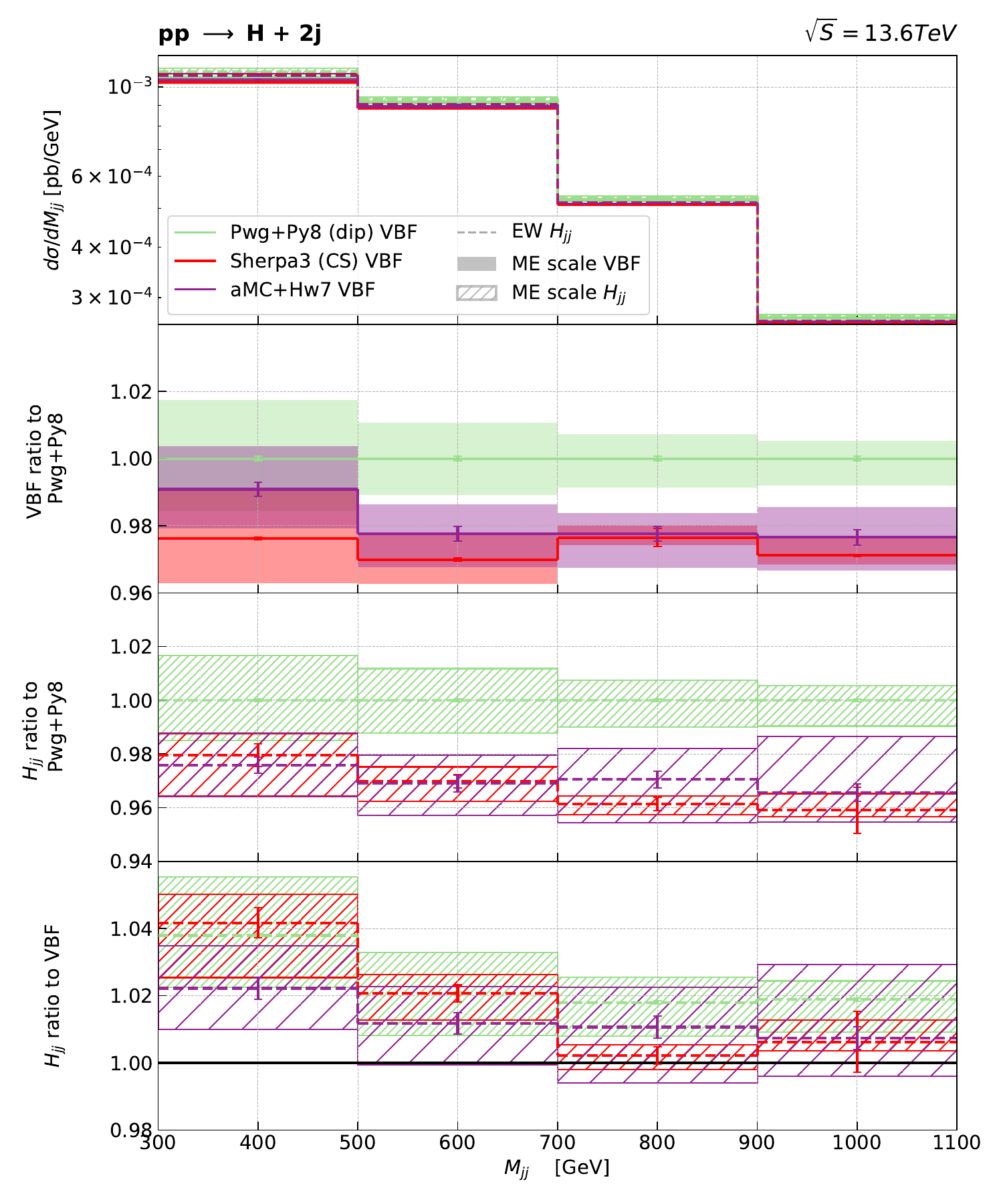}
  \caption{$4<|\Delta y_{\Pj\Pj}|<5$}
 \end{subfigure}
 \caption{
Dijet invariant mass distributions at NLOPS, considering the full EW $\PH\Pj\Pj$ production mode (dashed lines) and just the pure VBF contributions (solid lines) for a representative  \pythia{}~(green), \sherpa{}~(red), and \herwig{}~(violet) prediction,  for  two regions of dijet rapidity separation in the fiducial phase space \textbf{fiducial (b)}.
%
%
The solid (hatched) bands represent the uncertainty stemming from the 7-point scale variation in the hard matrix element calculations of the VBF (EW $\PH+2$ jets) production mode. 
The second (third) panel illustrates the ratio with respect to the central  \pythia{} prediction for the VBF ($\PH\Pj\Pj$) production mode.
The fourth panel illustrates the ratio between the $\PH\Pj\Pj$ predictions with respect to the central VBF one for each NLOPS generator. }
\label{fig:MJJ-DYJJ-FULL-vs-VBF}
\end{figure}  

In Sec.~\ref{sec:nlopscmp} we have considered QCD corrections to the pure VBF process.
It is, however, interesting to scrutinise the full EW $\PH$+2jets production (that we shorten as $\PH\Pj\Pj$), which comprises both the VBF and the $V\PH$ production modes.
For this study we choose three predictions, representative of the various parton showers and matching schemes, namely \PB{}+\pythia{} (simple shower with dipole recoil, in green), \madgraph{}+\herwig{}  (angular-ordered shower, in violet), and \sherpa{} (CS shower, in red). 

In particular, Fig.~\ref{fig:MJJ-DYJJ-FULL-vs-VBF} illustrates the dijet invariant mass within the same cuts as Fig.~\ref{fig:MJJ-DYJJ} for these three NLOPS predictions, including the uncertainty stemming from the renormalisation and factorisation scale variations in the hard matrix element, for the pure VBF contribution (solid line and band), and the full EW $\PH\Pj\Pj$ production (dashed lines and hatched bands).
As expected, for a moderate $\Delta y_{\Pj\Pj}$ separation (left plot),  the $\PH\Pj\Pj$ prediction is roughly $5-8\%$ larger the pure VBF one across our generators (ratio plot in the last panel), and uncertainties stemming from the hard matrix elements scale variations are roughly $4-5\%$.

Requiring $\Delta y_{\Pj\Pj}>4$~(right) strongly suppresses the $V\PH$ contribution, and the ratio between $\PH\Pj\Pj$ and  VBF is much closer to 1, with differences at most of $2\%$ in the lowest $M_{\Pj\Pj}$ bins.
However, while we notice that for the \PB{}+\pythia{} and \madgraph{}+\herwig{} generators, both the central values and the scale variations are similar between VBF and $\PH\Pj\Pj$, for \sherpa{} this is true only for the central value, while scale variations appear much larger.
As expected, the differences between the pure VBF sample and the full EW $\PH\Pj\Pj$ one are very similar across all the NLOPS samples.

\subsection{Considerations on parton shower and matching uncertainties}
\label{sec:NLOPSunc}

As discussed in Sec.~\ref{sec:expsummary}, parton shower and matching uncertainties amount to a sizeable contribution to the error budget of differential distributions.
Given the need for a coherent path for understanding and reducing such systematics, we provide here a recommendation for a robust assessment of parton shower and matching uncertainties, given the current state of the art.

To derive this recommendation we focus on the figures shown in Sec.~\ref{sec:nlopscmp}, but our considerations holds for all the distributions of the present study.

\subsubsection{Uncertainties from varying NLOPS generators}

From the middle panel of Figs.~\ref{fig:MJJ-PTH}, \ref{fig:MJJ-DYJJ}, and \ref{fig:PTH-NJETS}, it is evident that the error obtained by simply performing a scale variation on a single NLOPS sample (ME + PS starting scale) must be combined with the uncertainty arising from variations in matching and parton-shower schemes. 
It is also evident that the largest uncertainty stems from varying the parton shower, rather than the matching scheme.
This was already observed, \emph{e.g.}\ in Ref.~\cite{Buckley:2021gfw}.
In the bottom panels, we illustrate the breakdown of all the predictions contributing to this envelope. 
For inclusive distributions (such as those in Figs.~\ref{fig:MJJ-PTH}, \ref{fig:MJJ-DYJJ} and the left panel of Fig.~\ref{fig:PTH-NJETS}),  \herwig{} and \sherpa{} predictions are rather close to each other, while the \pythia{} showers always yield higher rates.
Looking at more exclusive observables (such as the right panel of Fig.~\ref{fig:PTH-NJETS}), \emph{i.e.}\ those that can be defined only if a third jet is present, 
again all the \sherpa{} and \herwig{} predictions appear quite close to each other.
Within the group of these neighbouring predictions, the most different is 
typically the \herwig{} angular-ordered shower with internal matching which has a higher rate by up to $10\%$.
Both \pythia{} curves instead undershoot the \sherpa{}/\herwig{} envelope by roughly 10\%.

In general, there is a remarkably good agreement between \herwig{} and \sherpa{}, regardless of the shower and the matching used, with at most 5\% discrepancies for inclusive observables, and at most 10\% differences for more exclusive ones.
\pythia predictions are somehow always outliers, with the default ``simple'' shower with dipole recoil and default hard scale setting being the closest to \herwig{} and \sherpa{}.

\subsubsection{Scale uncertainties in a NLOPS simulation}
 \begin{figure}[t!]
  \begin{subfigure}[t]{0.49\textwidth}
   \includegraphics[width=\textwidth]{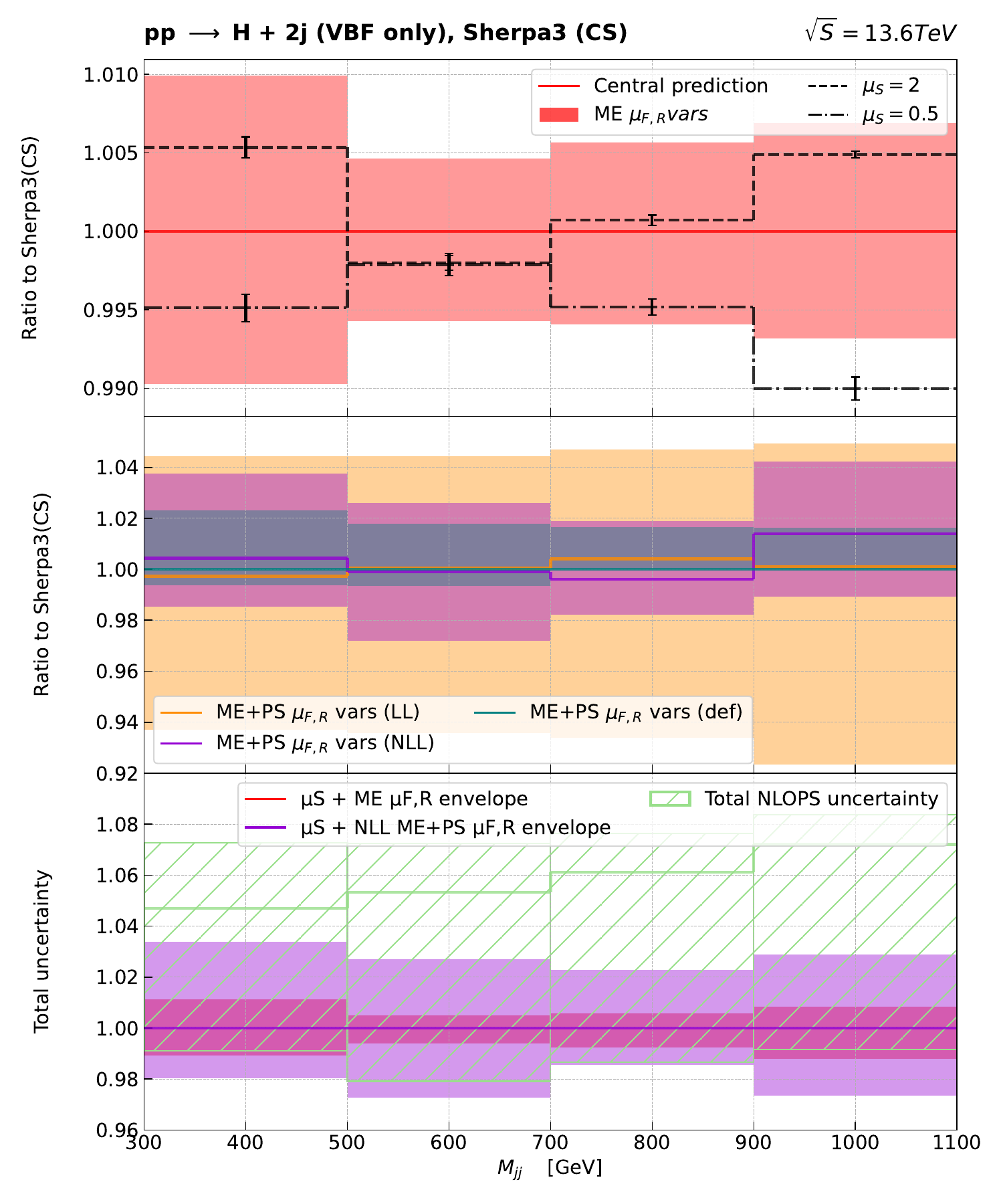}
  \caption{$2<|\Delta y_{\Pj\Pj}|<4$}
 \end{subfigure}%
  \begin{subfigure}[t]{0.49\textwidth}
 \includegraphics[width=\textwidth]{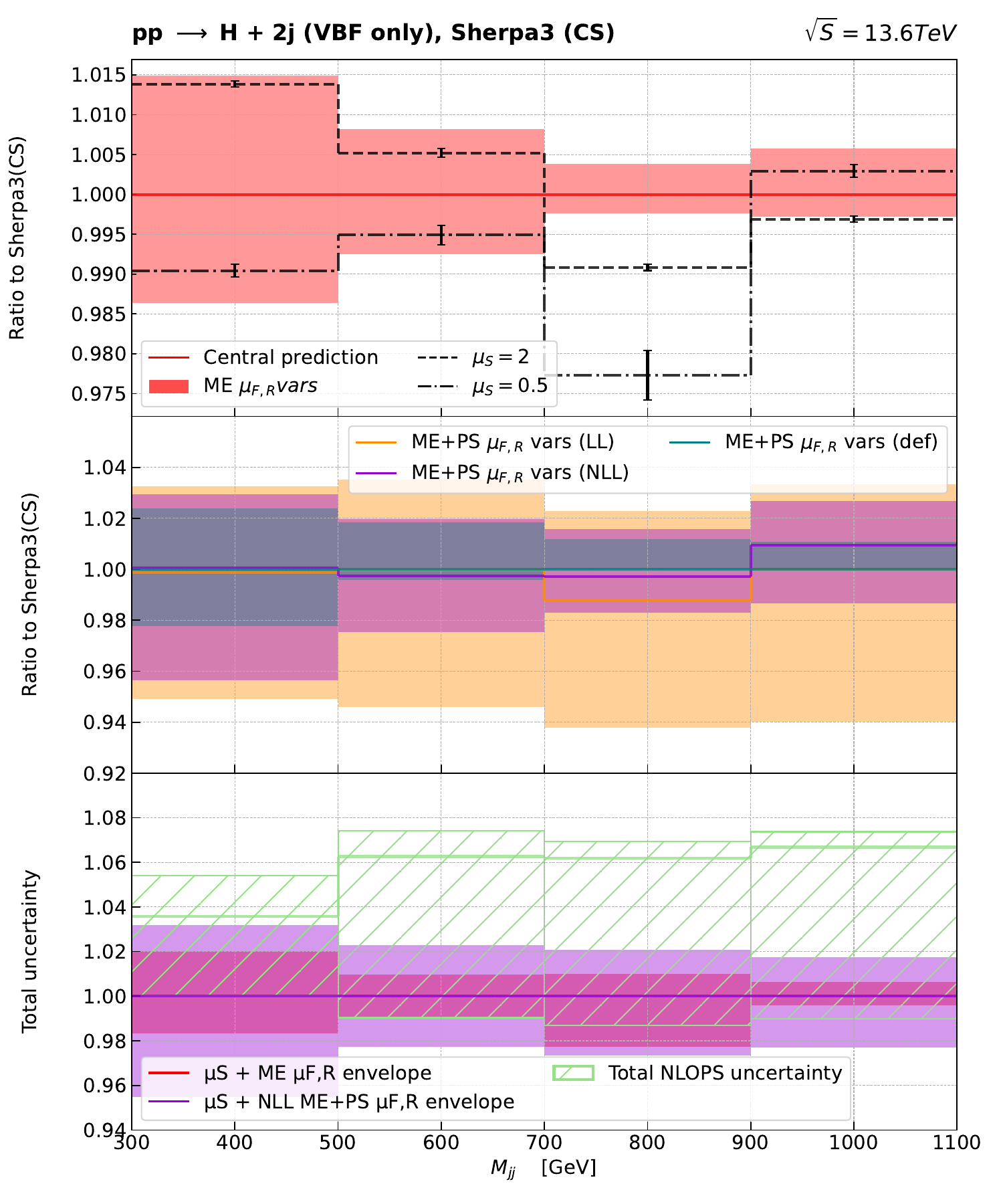}
  \caption{$4<|\Delta y_{\Pj\Pj}|<5$}
 \end{subfigure}
 \caption{
Relative uncertainty for the dijet mass distributions at NLOPS for the \sherpa{} default shower, breaking into two dijet rapidity separation $\Delta y_{\Pj\Pj}$ regions.
 Fiducial cuts are the same as in Fig.~\ref{fig:MJJ-DYJJ}.
}
\label{fig:MJJ-DYJJ-scales}
\end{figure}  

We now comment on scale variations in the parton showers.
For all generators used in this study, the structure of higher-order corrections to soft-gluon emission~\cite{Amati:1980ch} was used to deduce the functional form of the argument of the strong coupling. The renormalisation scale is therefore proportional to the transverse momentum of the emission, although the precise kinematical definition of the transverse momentum can differ among the codes. Variations of this scale are partially compensated by higher-order corrections that (at NLL) affect the soft emissions due to the combined structure of the leading poles and leading logarithms~\cite{Kodaira:1981nh,Davies:1984hs,Davies:1984sp,Catani:1988vd,Catani:1990rr,Dulat:2018vuy}.
The parton showers considered in this study generally have LL accuracy, with some NLL violating terms arising either from the inability to describe non-global observables (as is the case for the angular-ordered shower~\cite{Banfi:2006gy}) or from the  handling of multiple emissions at commensurate hardness (see \emph{e.g.}\ Refs.~\cite{vanBeekveld:2023chs,Dasgupta:2018nvj} for a detailed discussion).
However, all of the parton showers implement splitting functions compatible with NLL evolution, such that the NLL renormalisation scale variations can be used. The uncertainty stemming from NLL-violating terms is of kinematical origin and should be parametrised in different way, for example by varying the recoil scheme.

At NLL, the higher-order corrections proportional to the QCD beta function can be implemented as a term multiplying only soft emissions~\cite{Proceedings:2018jsb}. More specifically, if we wish to vary the argument of the running coupling by a factor $r$, the following functional form of the coupling should be used:
\begin{equation}
\alpha_s(r k_t) \to \alpha_s(r k_t)\left[ 1+ f(z)\frac{\alpha_s(r k_t)}{2\pi}\beta_0 \ln r^2\right],
\label{eq:muRvar}
\end{equation}
where $\beta_0$ is the one-loop QCD beta function, and $f(z)=1$ when the emission is soft, and $f(z)=0$ otherwise. 
\pythia is the only tool which implements this type of NLL-preserving scale variations in its current public version~\cite{Mrenna:2016sih}.
The \herwig{} shower implements Eq.~\eqref{eq:muRvar} with $f(z)=0$, consistent
with LL evolution but not with NLL.
The current public version of {\sc Sherpa}'s CS shower sets $f(z)=1$ by default, alternatively the LL correct $f(z)=0$ can be used.
For the purpose of this study, we have modified the \sherpa{} generator to also include NLL scale variations\footnote{This option will be part of \sherpa{} releases starting from version 3.1.}. 
We note that the {\sc Dire} parton shower use separate $\alpha_s$ schemes for the soft and collinear part of the splitting functions, and the {\sc Alaric} parton shower treats soft and collinear terms as separate splitting kernels~\cite{Herren:2022jej}.
In both cases the NLL scale variation is implemented.
In the case of {\sc Powheg} matching, we should also consider scale variations in the hardest emission generator.
However at present no variation of this kind is possible in the \PB, which only allows for renormalisation and factorisation scale variations entering the NLO normalisation of the event. 

Factorisation scale variations are currently possible only in \sherpa{} (on-the-fly) \cite{Bothmann:2016nao} and \herwig{}. They will become available in \pythia{} starting from release 8.316. 

The result for the $M_{\Pj\Pj}$ distribution, broken into two $\Delta y_{\Pj\Pj}$ ranges, is shown in Fig.~\ref{fig:MJJ-DYJJ-scales}.
All curves are normalised with respect to the central prediction (CS shower), so to better assess the size of the relative uncertainties.
In the first panel, we show the uncertainty from scale variations in the matrix element (red solid band) and from varying the shower starting scale of a factor 2 up (dashed black line) and down (dash dotted line).
In the second panel, we compare the band stemming from renormalisation and factorisation scale variations not only in the matrix element but also including a correlated variation in the shower splitting kernels, using the default option (renormalisation scale compensating term always present, in green), the one compatible with NLL evolution (renormalisation scale compensating term present only for soft emissions, as in \pythia{}, violet) and the one compatible with LL evolution (compensating term never present, as in \herwig{}, orange).
From the first panel, we see the scale variatiions in the pure matrix element is of the order of 1\% up and down, while the NLL variation is roughly 2-3 times bigger.
The default \sherpa{} variation obtained is of the order of 1-2\%, and thus often underestimates the NLL one, while the LL variation yields a much band with spread of around $10\%$.
In the bottom panel we sum in quadrature the uncertainty coming from the parton shower starting scale  together with the factorisation and renormalisation scale variation in the matrix element (red) or in the matrix element and in the shower using the NLL prescription for the renormalisation scale (violet).
The NLL (LL) band is smaller (comparable) in size with the one obtaining from the envelope of all NLOPS generators (bottom panel of Fig.~\ref{fig:MJJ-DYJJ}), which dominates the NLOPS uncertaintes as computed in the previous sections, represented by the green band in the bottom panel of Fig.~\ref{fig:MJJ-DYJJ-scales}.
Thus we believe that comparing different tools with the same formal accuracy has always to be performed.
However, in view of more accurate parton showers algorithms being developed, we encourage the authors of all the tools presented in this study to include an option for on-the-fly factorisation scale variation, and one for both LL and NLL renormalisation scale variations.

\subsubsection{Recommendations for NLOPS uncertainty estimates}
\label{sec:recommendations}

\begin{mdframed}
Based on these observations, we propose the following recommendation for a more robust estimation of NLOPS uncertainty, which comprises both the matching and the shower variations:
\begin{itemize}
\item[1.] Generate (at least) three predictions:
\begin{itemize}
 \item one with \PB{} and any \pythia{} shower;
 \item one with any \herwig{} shower;\footnote{Preferably with matching achieved through \textsc{Matchbox}, to ensure proper alignment of parameters between the hardest emission and parton shower generation.}
 \item one with any \sherpa{} shower.
\end{itemize}
Make sure that at least one of these prediction is based on {\sc Powheg} matching, and one on MC@NLO matching.\footnote{We remind the reader that the preferred matching scheme of \sherpa{}  is MC@NLO, for VBF in \pythia{} only matching via \PB{} is allowed, while  for \herwig{} all matching schemes are available.}
Then, construct an envelope\footnote{For inclusive distributions, since \herwig{} and \sherpa{} yield similar results, one of these curves can be omitted. However, we recommend verifying this similarity for more exclusive observables.} out of these predictions.
\item[2.] For one of the curves computed in step~1:
\begin{itemize}
\item compute a shower starting scale variation band (or a \texttt{pThard} variation in the context of \PB+\pythia matching); 
\item compute the renormalisation and factorisation scale variation band, possibly including also scale variations in the shower.
\end{itemize}
Sum these bands in quadrature and combine this uncertainty in quadrature with the first envelope uncertainty.
\end{itemize}
\end{mdframed}
Notice that, contrary to what is done in many experimental analyses, we do not distinguish between matching and parton-shower uncertainties, but rather provide a unified band, because the two aspects are often correlated, \emph{i.e.}\ a matching procedure is usually tailored to the parton shower it is applied to.

We want to remind the reader that our recommendation is based on the fact that all currently available showers that can be matched with NLO QCD calculations for VBF have the same leading-logarithmic accuracy\footnote{An exception is the default \pythia{} shower with global recoil, which violates color coherence and should be avoided. For this reason this shower is not considered in this study.}. 
Consequently, the predictions are equivalent, and their spread provides a genuine measure of the systematic error.
This approach may need revision when NLL showers matched to NLO calculations become publicly available in fully-fledged GPMC generators.
This would also imply the use of NLL scale variations.

We finally emphasise that the present study addresses only perturbative uncertainties. However,  comparing three different GPMC generators — \pythia{}, \sherpa{}, and \herwig{} — which implement distinct soft physics models (\emph{e.g.}\ hadronisation and multi-parton interactions) while aligning all input parameters related to the hard processes as done here, should offer a reasonably robust assessment of non-perturbative uncertainties as well.
A more detailed analysis of such uncertainties lies beyond the scope of this work.

\section{Summary}
\label{sec:summary}

The production of Higgs bosons via vector-boson fusion (VBF) is one of the primary Higgs-boson production mechanisms at the LHC.
In this context, current investigations of Higgs-boson physics necessitate a deeper exploration of this channel.
Precise and reliable theoretical predictions are therefore essential to fully exploit the potential of experimental data.

In this work, we first reviewed past experimental and theoretical studies on VBF at the LHC.
This summary, particularly the accompanying tables, aims to facilitate the proper citation of VBF studies, especially of theoretical work.
Additionally, we have provided theoretical predictions at fixed order and with parton-shower corrections in generic fiducial phase space regions, as well as within the STXS framework.

Regarding fixed-order predictions, we have discussed in detail all contributions to the process $\process$.
The contributions concerning VBF production have been calculated with state-of-the-art accuracy while the background contributions have been provided for reference only and hence with lower accuracy.
The results are presented in the form of cross sections and differential distributions.

We performed a detailed comparison of different state-of-the-art NLOPS predictions, for both the VBF channel as well as the full EW $\PH\Pj\Pj$ production.
This enabled us to formulate a prescription for assessing the theory uncertaintes: one should consider three-points variation (\sherpa, \pythia, \herwig) and include consistently factorisation-, renormalisation-, and starting-scale variations. 
While we focused mostly on the VBF fiducial phase space region, 
we believe similar prescriptions can be a solid way of assessing NLOPS uncertainties for more generic processes.

With this work, we have gained a 
better understanding of current theory uncertainties for VBF production at the LHC bearing in mind the current available theoretical calculations.
These findings are expected to be of great use during the run~III of the LHC, where dedicated VBF measurements are foreseen.
Since in our study we mainly focused on the signal process, 
we believe a similar collective effort is also required to get a clearer understanding of background contamination in VBF phase space regions with state-of-the-art predictions~\cite{Chen:2025whf}.

Finally, we recall that all the results and auxiliary data files can be publicly accessed at:
\begin{center}
\url{https://gitlab.cern.ch/LHCHIGGSXS/LHCHXSWG1/VBFStudyYR5}
\end{center}
We hope that this work will be useful to the community, ease the acknowledgement of theoretical contributions as well as enable the optimal use of state-of-the-art theoretical predictions in experimental analyses.
In doing so, it will support further investigations of Higgs boson production via the VBF channel at the LHC.

\section*{Acknowledgements}

The authors are grateful to the LHC Higgs Cross Section Working Group for providing such a framework and support for the present work.
We want to thank Joey Huston and Simon Pl\"atzer for useful discussions.
This work has also greatly benefited from the PhysTeV workshop 2023 at Les Houches, France~\cite{Andersen:2024czj}.

This research was supported by FermiForward Discovery Group, LLC
under Contract No. 89243024CSC000002 with the U.S. Department of Energy,
Office of Science, Office of High Energy Physics.
BJ, MP, and SR acknowledge support by the state of Baden-Württemberg through bwHPC
and the German Research Foundation (DFG) through grant No.~INST 39/963-1 FUGG (bwForCluster NEMO).
MP acknowledges support by the DFG through the Research Training Group RTG2044.
CTP thanks the PLEIADES HPC team at the University of Wuppertal,
supported by the DFG through grant No.~INST 218/78-1 FUGG and the Bundesministerium
f\"ur Bildung und Forschung (BMBF).
DR acknowledges support by the European Union under the HORIZON program in Marie
Sk{\l}odowska-Curie project No.\ 101153541.
MZ~acknowledges the financial support by the MUR (Italy), with
funds of the European Union (NextGenerationEU), through the PRIN2022
grant 2022EZ3S3F.

\bibliography{vbf-higgs-wg}

\end{document}